\documentclass[leqno,epsf,12pt]{amsart}
\input epsf.sty
\usepackage{amsmath,epsfig}
\usepackage{amssymb}
\usepackage{amsfonts,amsthm,amscd}
\usepackage{hyperref}
\def\p{\;\raisebox{-1.5mm}{\epsfysize=6mm\epsfbox{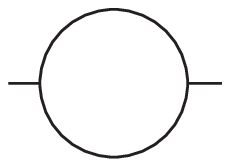}}\;}
\def\pdp{\;\raisebox{-1.5mm}{\epsfysize=6mm\epsfbox{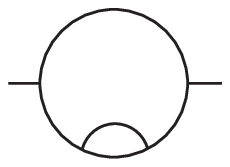}}\;}
\def\pdpdp{\;\raisebox{-1.5mm}{\epsfysize=6mm\epsfbox{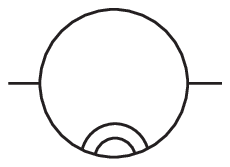}}\;}
\def\pdpdpdp{\;\raisebox{-1.5mm}{\epsfysize=6mm\epsfbox{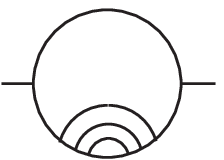}}\;}
\def\pddpp{\;\raisebox{-1.5mm}{\epsfysize=6mm\epsfbox{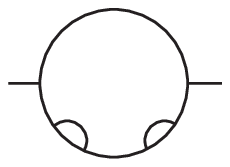}}\;}
\numberwithin{equation}{section}

\newtheorem{theorem}{Theorem}[section]
\newtheorem{corollary}[theorem]{Corollary}
\newtheorem{lemma}[theorem]{Lemma}
\newtheorem{proposition}[theorem]{Proposition}
\theoremstyle{definition}
\newtheorem{defn}[theorem]{Definition}
\newtheorem*{acknowledgments}{Acknowledgments}
\theoremstyle{remark}
\newtheorem{remark}[theorem]{Remark}

\newcommand{\nc}{\newcommand}
\newcommand{\be}{\begin{equation}}
\newcommand{\ee}{\end{equation}}

\newcommand{\bc}{\begin{center}}
\newcommand{\ec}{\end{center}}
\nc{\bth}{\begin{theorem}} \nc{\bpr}{\begin{proposition}}
\nc{\epr}{\end{proposition}} \nc{\ble}{\begin{lemma}}
\nc{\ele}{\end{lemma}} \nc{\bco}{\begin{corollary}}
\nc{\eco}{\end{corollary}} \nc{\bre}{\begin{remark}}
\nc{\ere}{\end{remark}}
   \nc{\f}{\frac}
   \nc{\pa}{\partial}
   \nc{\na}{\nabla}
   \nc{\al}{\alpha}
   \nc{\bet}{\beta}
   \nc{\ga}{\gamma}
   \nc{\de}{\delta}
   \nc{\De}{\Delta}
   \nc{\Om}{\Omega}
   \nc{\om}{\omega}
   \nc{\tom}{\tilde\omega}
   \nc{\ep}{\varepsilon}
   \nc{\tvp}{\tilde\varphi}
   \nc{\vp}{\varphi}
   \nc{\R}{\mathbb R}
   \nc{\Z}{\mathbb Z}
   \nc{\CC}{\mathbb C}
   \nc{\Ad}{\rm Ad}
   \nc{\fraka}{{\mathfrak g}_{\mathcal A}}
   \nc{\Ga}{G_{\mathcal A}}

\usepackage[all]{xy}               
\usepackage{exscale}
\usepackage{color}
\usepackage{axodraw}
\usepackage{colordvi}

\def\at1{\begin{array}{c} \ta1\ \\ \end{array}}
\def\Mat31{\begin{array}{c} \td31\ \\ \end{array}}
\def\mat41{\begin{array}{c} \tb2\ \\ \end{array}}
\def\mot43{\begin{array}{c} \th43\ \\ \end{array}}

\def\ta1{{\scalebox{0.25}{ 
\begin{picture}(12,12)(38,-38)
\SetWidth{0.5} \SetColor{Black} \Vertex(45,-33){5.66}
\end{picture}}}}

\def\tb2{{\scalebox{0.25}{ 
\begin{picture}(12,42)(38,-38)
\SetWidth{0.5} \SetColor{Black} \Vertex(45,-3){5.66}
\SetWidth{1.0} \Line(45,-3)(45,-33) \SetWidth{0.5}
\Vertex(45,-33){5.66}
\end{picture}}}}

\def\tc3{{\scalebox{0.25}{ 
\begin{picture}(12,72)(38,-38)
\SetWidth{0.5} \SetColor{Black} \Vertex(45,27){5.66}
\SetWidth{1.0} \Line(45,27)(45,-3) \SetWidth{0.5}
\Vertex(45,-33){5.66} \SetWidth{1.0} \Line(45,-3)(45,-33)
\SetWidth{0.5} \Vertex(45,-3){5.66}
\end{picture}}}}

\def\td31{{\scalebox{0.25}{ 
\begin{picture}(42,42)(23,-38)
\SetWidth{0.5} \SetColor{Black} \Vertex(45,-3){5.66}
\Vertex(30,-33){5.66} \Vertex(60,-33){5.66} \SetWidth{1.0}
\Line(45,-3)(30,-33) \Line(60,-33)(45,-3)
\end{picture}}}}

\def\te4{{\scalebox{0.25}{ 
\begin{picture}(12,102)(38,-8)
\SetWidth{0.5} \SetColor{Black} \Vertex(45,57){5.66}
\Vertex(45,-3){5.66} \Vertex(45,27){5.66} \Vertex(45,87){5.66}
\SetWidth{1.0} \Line(45,57)(45,27) \Line(45,-3)(45,27)
\Line(45,57)(45,87)
\end{picture}}}}

\def\tf41{{\scalebox{0.25}{ 
\begin{picture}(42,72)(38,-8)
\SetWidth{0.5} \SetColor{Black} \Vertex(45,27){5.66}
\Vertex(45,-3){5.66} \SetWidth{1.0} \Line(45,27)(45,-3)
\SetWidth{0.5} \Vertex(60,57){5.66} \SetWidth{1.0}
\Line(45,27)(60,57) \SetWidth{0.5} \Vertex(75,27){5.66}
\SetWidth{1.0} \Line(75,27)(60,57)
\end{picture}}}}

\def\tg42{{\scalebox{0.25}{ 
\begin{picture}(42,72)(8,-8)
\SetWidth{0.5} \SetColor{Black} \Vertex(45,27){5.66}
\Vertex(45,-3){5.66} \SetWidth{1.0} \Line(45,27)(45,-3)
\SetWidth{0.5} \Vertex(15,27){5.66} \Vertex(30,57){5.66}
\SetWidth{1.0} \Line(15,27)(30,57) \Line(45,27)(30,57)
\end{picture}}}}

\def\th43{{\scalebox{0.25}{ 
\begin{picture}(42,42)(8,-8)
\SetWidth{0.5} \SetColor{Black} \Vertex(45,-3){5.66}
\Vertex(15,-3){5.66} \Vertex(30,27){5.66} \SetWidth{1.0}
\Line(15,-3)(30,27) \Line(45,-3)(30,27) \Line(30,27)(30,-3)
\SetWidth{0.5} \Vertex(30,-3){5.66}
\end{picture}}}}

\def\thj44{{\scalebox{0.25}{ 
\begin{picture}(42,72)(8,-8)
\SetWidth{0.5} \SetColor{Black} \Vertex(30,57){5.66}
\SetWidth{1.0} \Line(30,57)(30,27) \SetWidth{0.5}
\Vertex(30,27){5.66} \SetWidth{1.0} \Line(45,-3)(30,27)
\SetWidth{0.5} \Vertex(45,-3){5.66} \Vertex(15,-3){5.66}
\SetWidth{1.0} \Line(15,-3)(30,27)
\end{picture}}}}

\def\ti5{{\scalebox{0.25}{ 
\begin{picture}(12,132)(23,-8)
\SetWidth{0.5} \SetColor{Black} \Vertex(30,117){5.66}
\SetWidth{1.0} \Line(30,117)(30,87) \SetWidth{0.5}
\Vertex(30,87){5.66} \Vertex(30,57){5.66} \Vertex(30,27){5.66}
\Vertex(30,-3){5.66} \SetWidth{1.0} \Line(30,-3)(30,27)
\Line(30,27)(30,57) \Line(30,87)(30,57)
\end{picture}}}}

\def\tj51{{\scalebox{0.25}{ 
\begin{picture}(42,102)(53,-38)
\SetWidth{0.5} \SetColor{Black} \Vertex(61,27){4.24}
\SetWidth{1.0} \Line(75,57)(90,27) \Line(60,27)(75,57)
\SetWidth{0.5} \Vertex(90,-3){5.66} \Vertex(60,27){5.66}
\Vertex(75,57){5.66} \Vertex(90,-33){5.66} \SetWidth{1.0}
\Line(90,-33)(90,-3) \Line(90,-3)(90,27) \SetWidth{0.5}
\Vertex(90,27){5.66}
\end{picture}}}}

\def\tk52{{\scalebox{0.25}{ 
\begin{picture}(42,102)(23,-8)
\SetWidth{0.5} \SetColor{Black} \Vertex(60,57){5.66}
\Vertex(45,87){5.66} \SetWidth{1.0} \Line(45,87)(60,57)
\SetWidth{0.5} \Vertex(30,57){5.66} \SetWidth{1.0}
\Line(30,57)(45,87) \SetWidth{0.5} \Vertex(30,-3){5.66}
\SetWidth{1.0} \Line(30,-3)(30,27) \SetWidth{0.5}
\Vertex(30,27){5.66} \SetWidth{1.0} \Line(30,57)(30,27)
\end{picture}}}}

\def\tl53{{\scalebox{0.25}{ 
\begin{picture}(42,102)(8,-8)
\SetWidth{0.5} \SetColor{Black} \Vertex(30,57){5.66}
\Vertex(30,27){5.66} \SetWidth{1.0} \Line(30,57)(30,27)
\SetWidth{0.5} \Vertex(30,87){5.66} \SetWidth{1.0}
\Line(30,27)(45,-3) \SetWidth{0.5} \Vertex(15,-3){5.66}
\SetWidth{1.0} \Line(15,-3)(30,27) \Line(30,57)(30,87)
\SetWidth{0.5} \Vertex(45,-3){5.66}
\end{picture}}}}

\def\tm54{{\scalebox{0.25}{ 
\begin{picture}(42,72)(8,-38)
\SetWidth{0.5} \SetColor{Black} \Vertex(30,-3){5.66}
\SetWidth{1.0} \Line(30,27)(30,-3) \Line(30,-3)(45,-33)
\SetWidth{0.5} \Vertex(15,-33){5.66} \SetWidth{1.0}
\Line(15,-33)(30,-3) \SetWidth{0.5} \Vertex(45,-33){5.66}
\SetWidth{1.0} \Line(30,-33)(30,-3) \SetWidth{0.5}
\Vertex(30,-33){5.66} \Vertex(30,27){5.66}
\end{picture}}}}

\def\tn55{{\scalebox{0.25}{ 
\begin{picture}(42,72)(8,-38)
\SetWidth{0.5} \SetColor{Black} \Vertex(15,-33){5.66}
\Vertex(45,-33){5.66} \Vertex(30,27){5.66} \SetWidth{1.0}
\Line(45,-33)(45,-3) \SetWidth{0.5} \Vertex(45,-3){5.66}
\Vertex(15,-3){5.66} \SetWidth{1.0} \Line(30,27)(45,-3)
\Line(15,-3)(30,27) \Line(15,-3)(15,-33)
\end{picture}}}}

\def\tp56{{\scalebox{0.25}{ 
\begin{picture}(66,111)(0,0)
\SetWidth{0.5} \SetColor{Black} \Vertex(30,66){5.66}
\Vertex(45,36){5.66} \SetWidth{1.0} \Line(30,66)(45,36)
\Line(15,36)(30,66) \SetWidth{0.5} \Vertex(30,6){5.66}
\Vertex(60,6){5.66} \SetWidth{1.0} \Line(60,6)(45,36)
\SetWidth{0.5}
\SetWidth{1.0} \Line(45,36)(30,6) \SetWidth{0.5}
\Vertex(15,36){5.66}
\end{picture}}}}

\def\tq57{{\scalebox{0.25}{ 
\begin{picture}(81,111)(0,0)
\SetWidth{0.5} \SetColor{Black} \Vertex(45,36){5.66}
\Vertex(30,6){5.66} \Vertex(60,6){5.66} \SetWidth{1.0}
\Line(60,6)(45,36) \SetWidth{0.5}
\SetWidth{1.0} \Line(45,36)(30,6) \SetWidth{0.5}
\Vertex(75,36){5.66} \SetWidth{1.0} \Line(45,36)(60,66)
\Line(60,66)(75,36) \SetWidth{0.5} \Vertex(60,66){5.66}
\end{picture}}}}

\def\tr58{{\scalebox{0.25}{ 
\begin{picture}(81,111)(0,0)
\SetWidth{0.5} \SetColor{Black} \Vertex(60,6){5.66}
\Vertex(75,36){5.66} \SetWidth{1.0} \Line(60,66)(75,36)
\SetWidth{0.5} \Vertex(60,66){5.66}
\SetWidth{1.0} \Line(60,36)(60,66) \Line(60,6)(60,36)
\SetWidth{0.5} \Vertex(60,36){5.66} \Vertex(45,36){5.66}
\SetWidth{1.0} \Line(60,66)(45,36)
\end{picture}}}}

\def\ts59{{\scalebox{0.25}{ 
\begin{picture}(81,111)(0,0)
\SetWidth{0.5} \SetColor{Black}
\Vertex(75,36){5.66} \SetWidth{1.0} \Line(60,66)(75,36)
\SetWidth{0.5} \Vertex(60,66){5.66}
\SetWidth{1.0} \Line(60,36)(60,66) \SetWidth{0.5}
\Vertex(60,36){5.66} \Vertex(45,36){5.66} \SetWidth{1.0}
\Line(60,66)(45,36) \Line(75,6)(75,36) \SetWidth{0.5}
\Vertex(75,6){5.66}
\end{picture}}}}

\def\tz591{{\scalebox{0.25}{ 
\begin{picture}(81,111)(0,0)
\SetWidth{0.5} \SetColor{Black}
\Vertex(75,36){5.66} \SetWidth{1.0} \Line(60,66)(75,36)
\SetWidth{0.5} \Vertex(60,66){5.66}
\SetWidth{1.0} \Line(60,36)(60,66) \SetWidth{0.5}
\Vertex(60,36){5.66} \Vertex(45,36){5.66} \SetWidth{1.0}
\Line(60,66)(45,36) \SetWidth{0.5} \Vertex(45,6){5.66}
\SetWidth{1.0} \Line(45,6)(45,36)
\end{picture}}}}

\begin{document}
\normalsize
\title[Feynman diagrams and Lax pair equations]{Feynman diagrams and Lax pair equations}
\date{\today. Erwin Schr\"odinger Institut preprint number 2144. AMS classification: 81T15,17B80.  Keywords: Renormalization,
 Lax pair equations, Hopf algebras.  }
\author{Gabriel B\u adi\c toiu}
\address{ Institute of Mathematics of the Romanian
   Academy, PO Box 1-764, 014700 Bucharest, Romania and
   Max-Planck-Institut f\" ur Mathematik, P.O. Box 7280, D-53072 Bonn,
   Germany.
  \texttt{baditoiu@math.bu.edu}}
\author{Steven Rosenberg}
\address{Department of Mathematics and Statistics, Boston
   University, Boston, MA 02215, USA. \texttt{sr@math.bu.edu}}

\begin{abstract}
   We find a Lax pair equation corresponding to the Connes-Kreimer
   Birkhoff factorization of the character group of a Hopf algebra.  This flow
   preserves the locality of counterterms. 
 In particular, we obtain a flow for the character
   given by Feynman rules, and relate this flow to the Renormalization Group
   Flow. \end{abstract}
\maketitle

\section{Introduction}
  In the theory of integrable systems, many classical mechanical systems
  are described by
a Lax pair
  equation associated to a coadjoint orbit of a semisimple Lie group, for
  example via 
the Adler-Kostant-Symes theorem \cite{adler}. Solutions are given 
 by a Birkhoff factorization on the 
group, and in some cases, this technique extends to loop group formulations of
 physically interesting systems such as the Toda lattice
\cite{guest, sts}. By the work of Connes-Kreimer \cite{ck1}, there is a
Birkhoff factorization of characters on general Hopf algebras, in particular
on the Kreimer Hopf
  algebra  of 1PI Feynman diagrams.
In this paper, we reverse the usual procedure in
  integrable systems: we
construct  a Lax pair equation $\frac{d L}{dt}=[L,M]$ 
on the Lie algebra of
 infinitesimal characters of the Hopf algebra 
whose solution is
  given precisely by the Connes-Kreimer Birkhoff factorization
  (Theorem~\ref{t:8.2}). The Lax pair equation is nontrivial in the sense that
it is not an infinitesimal inner automorphism.  
The main
  technical issue, that the Lie algebra of infinitesimal characters is not
  semisimple, is overcome by passing to the double Lie algebra with
  the simplest possible Lie algebra structure.  In particular,
  the Lax pair equation induces a flow for the character  given by
  Feynman rules in dimensional regularization.  This flow
 has the physical
  significance that it preserves locality, the independence of the character's counterterm
 on the mass parameter.

In \S\S1-4, we introduce a method to produce a Lax pair on any Lie
algebra from equations of motion on the double Lie algebra. In
\S\ref{hopf}, we apply this method to the particular case of the Lie
algebra of infinitesimal characters of a Hopf algebra, and
prove Theorem~\ref{t:8.2}.

The
Renormalization Group Flow (RGF) usually considered in quantum field theory is
a flow on the character group $G_{\mathcal A}$, while the Lax pair flow is on the corresponding
Lie algebra ${\mathfrak g}_{\mathcal A}$ of infinitesimal characters. 
 There are various bijections from ${\mathfrak g}_{\mathcal A}$ to
 $G_{\mathcal A}$,
 and via these bijections we can compare the Lax pair
flow to the RGF. These flows are not the same, so we study how physically
significant quantities behave under the Lax pair flow.
In \S6, we derive an equation for the flow of
the $\beta$-function of  characters $\varphi_t\in G_{\mathcal A}$
associated to the Lax pair flow via the exponential map
$\exp: \mathfrak
g_{\mathcal A}\to G_{\mathcal A}$
(Corollary~\ref{cor:beta}).
 In \S7, we first 
show that the
Lax pair flow is trivial on primitives in the Hopf algebra.
We then use
 Manchon's bijection \cite{man} $\tilde R^{-1}:\mathfrak
g_{\mathcal A}\to G_{\mathcal A}$ 
to  prove various
locality results (Theorems~\ref{t:7.9}, \ref{t:7.14}). 
The $\beta$-function flow defined via $\tilde R^{-1}$ itself satisfies a Lax
pair equation (Theorem \ref{t:corr}).  Thus
 $\tilde R^{-1}$ is much better behaved than the exponential map.  In \S8, we work out several
examples of this theory, and in particular keep track of the leading log terms.  

An alternative algebraic geometric approach to
Lax pair equations is to apply
 spectral curve techniques to linearize
  the flow on the Jacobian of the spectral curve. Unfortunately, in the worked
  example of \S\ref{worked-example}, the spectral curve is  reducible, and the only
  invariants we find are trivial.  We hope to find examples with nontrivial
  invariants in the future.

We would like to thank Dirk Kreimer for suggesting we investigate the
connection between the
Connes-Kreimer factorization and integrable systems, and Dominique Manchon for
helpful conversations.

\section{The double Lie algebra and its associated Lie Group}

There is a well known method to associate a Lax pair equation to a
Casimir element on the dual $\mathfrak g^*$ of a semisimple Lie
algebra $\mathfrak g$ \cite{sts}. The semisimplicity is used to
produce an $\mathrm{Ad}$-invariant, symmetric, non-degenerate
bilinear form on $\mathfrak g$, allowing an identification of
$\mathfrak g$ with $\mathfrak g^*$.
  For a general Lie algebra  $\mathfrak g$,
there may be no such bilinear form. To produce a Lax pair, we need
to extend  $\mathfrak g$ to a larger Lie algebra with the desired
bilinear form. We do this by constructing a Lie bialgebra structure
on  $\mathfrak g$, whose definition we now recall (see
e.g.~\cite{ksch}).
\begin{defn}\label{bialg}
  A Lie bialgebra is a Lie algebra $(\mathfrak g, [\cdot ,\cdot ])$
  with a linear map
  $\gamma:\mathfrak g\to\mathfrak g\otimes\mathfrak g$ such that
  \begin{itemize}
     \item[a)]
         $^t\gamma:\mathfrak g^*\otimes\mathfrak g^*\to\mathfrak g^*$ defines
         a Lie bracket on $\mathfrak g^*$,
     \item[b)]
         $\gamma$ is a $1$-cocycle of $\mathfrak g$, i.e.
         $$\mathrm{ad}^{(2)}_x(\gamma(y))-\mathrm{ad}^{(2)}_y(\gamma(x))-\gamma([x,y])=0,$$
         where $\mathrm{ad}^{(2)}_x:\mathfrak g\otimes\mathfrak
         g\to\mathfrak g\otimes\mathfrak g$ is given by
         $\mathrm{ad}^{(2)}_x(y\otimes z)=\mathrm{ad}_x(y)\otimes z+y\otimes
         \mathrm{ad}_x(z)
= [x,y]\otimes z + y \otimes [x,z]$.
  \end{itemize}
\end{defn}

  A Lie bialgebra $(\mathfrak g,[\cdot,\cdot ],\gamma)$ induces an Lie
  algebra structure on the {\it double Lie algebra}
$\mathfrak g\oplus\mathfrak g^*$ by
  $$[X,Y]_{\mathfrak g\oplus\mathfrak g^*}=[X,Y],$$
  $$[X^*,Y^*]_{\mathfrak g\oplus\mathfrak g^*}= {}^t\gamma(X\otimes Y),$$
  $$[X,Y^*]=\mathrm{ad}^*_X(Y^*),$$
  for  $X$, $Y\in\mathfrak g$ and $X^*$, $Y^*\in\mathfrak g^*$,
  where $\mathrm{ad}^*$ is the coadjoint representation given by
  $\mathrm{ad}^*_X(Y^*)(Z)=-Y^*(\mathrm{ad}_X(Z))$ for
  $Z\in\mathfrak g$.

Since it is difficult to construct explicitly the Lie group associated to
the Lie algebra $\mathfrak g\oplus\mathfrak g^*$, we will
choose the trivial Lie bialgebra given by the cocycle $
\gamma=0$ and
denote by $\delta=\mathfrak g\oplus \mathfrak g^*$ the associated Lie
algebra. Let $\{Y_i,
i= 1,\ldots ,l\}$
be a basis of $\mathfrak g$, with dual basis
 $\{Y^*_i\}$.
The Lie bracket $[ \cdot,\cdot ]_\delta$  on $\delta$ is given by
 $$[Y_i,Y_j]_\delta=[Y_i,Y_j],\ [Y_i^*,Y_j^*]_\delta=0,\
   [Y_i,Y_j^*]_\delta=-\sum_k c^j_{ik}Y^*_k,$$
where the $c^j_{ik}$ are the structure constants:
$[Y_i,Y_j]=\sum_kc^k_{ij}Y_k$.
The Lie group naturally associated to $\delta$ is given by the
following proposition.

\bpr\label{prop22}
   Let $G$ be the simply connected Lie group with Lie algebra
   $\mathfrak{g}$
   and let $\theta:G\times \mathfrak g^*\to \mathfrak g^*$ be
   the coadjoint representation
   $\theta(g,X)=\mathrm{Ad}^*_{G}(g)(X)$. Then the Lie algebra of the
   semi-direct product $\tilde G =
G\ltimes_\theta \mathfrak g^*$ is the double
   Lie algebra $\delta$.
\epr

\begin{proof}
  The Lie group law on  the semi-direct product $\tilde
  G$
is given by
  $$(g,X)\cdot (g',X')=(gg',X+\theta(g,X')).$$
  Let $\tilde{\mathfrak g}$ be the Lie algebra of $\tilde G$. Then the
  bracket on $\tilde{\mathfrak g}$  is given by
  $$
   [X,Y^*]_{\tilde{ \mathfrak g}}
    =d\theta(X,Y^*), \ \ [X,Y]_{\tilde {\mathfrak
    g}}=[X,Y],\ \  [X^*,Y^*]_{\tilde {\mathfrak g}}=0,
   $$
   for left-invariant vector fields $X$,
  $Y$ of $G$ and  $X^*, Y^*\in\mathfrak g^*$. We have
  $d\theta(X,Y^*)=d\mathrm{Ad}^*_{G}(X)(Y^*)=[X,Y^*]_\delta$ since
  $d\mathrm{Ad}_{G}=\mathrm{ad}_{\mathfrak g}$.
\end{proof}

The main point of this construction is  existence of a good bilinear form on
the double.

\begin{lemma}
The natural pairing
 $\langle\cdot, \cdot\rangle:\delta\otimes\delta\to{\mathbb C}$ given by
 $$\langle (a, b^*), (c,d^*)\rangle = d^*(a) + b^*(c), \ \ \ a,c\in\mathfrak{g},\ \
 b^*, d^*\in\mathfrak{g^*},$$ is an
$\mathrm{Ad}$-invariant symmetric non-degenerate bilinear form on
the Lie algebra $\delta$.
\end{lemma}

\begin{proof} By \cite{ksch}, this bilinear form is ad-invariant.  Since
 $\tilde G$ is simply connected, the Ad-invariance follows.  As an explicit
 example, we have
$$\mathrm{Ad}_{\tilde G}((g,0))(Y_i,0) = (\mathrm{Ad}_G(g)(Y_i),0),\ \
 \text{and}\ \  \mathrm{Ad}_{\tilde 
G}((g,0))(0,Y_j^*) = (0, \mathrm{Ad}_{G}^*(g)(Y_j^*)),
$$
from which the invariance under $\Ad_{\tilde G}(g,0)$ follows.
\end{proof}

\section{The loop algebra of a Lie algebra}
Following \cite{adler}, we consider the loop algebra
$$L\delta=\{L(\lambda)=\sum\limits_{j=M}^N \lambda^jL_j \ |
                    \ M,N\in \mathbb Z, L_j\in\delta\}.$$
The natural Lie bracket on $L\delta $ is given by
$$\left[\sum \lambda^iL_i,\sum \lambda^j L_j'\right]=
\sum\limits_k \lambda^k\sum\limits_{i+j=k}[L_i,L_j'].$$
Set
\begin{eqnarray*}L\delta _+ &=&
\{L(\lambda)= \sum\limits_{j=0}^N
\lambda^jL_j \ | \ N\in\mathbb Z^+\cup \{0\},  L_j\in\delta\}\\
L\delta _-&=&
\{L(\lambda)=\sum\limits_{j=-M}^{-1}   \lambda^jL_j \ | \ M\in\mathbb Z^+,
L_j\in\delta\}.
\end{eqnarray*}
Let $P_+:L\delta \to L\delta _+$ and $P_-:L\delta \to L\delta _-$ be
the natural projections and set $R=P_+-P_-$.

The natural pairing
$\langle\cdot,\cdot\rangle$ on $\delta$
yields an $\mathrm{Ad}$-invariant, symmetric,
non-degenerate pairing on $L\delta $ by setting
   $$\left\langle\sum\limits_{i=M}^N \lambda^iL_i ,\sum\limits_{j=M'}^{N'}
     \lambda^jL'_j \right\rangle=
     \sum\limits_{i+j=-1}\langle L_i,L_j'\rangle.
   $$

For our choice of basis $\{Y_i\}$ of $\mathfrak g$, we get an isomorphism
\begin{equation}\label{I}I: L(\delta ^*)\to L\delta \end{equation}
   with
      $$I\left(\sum L^j_iY_j\lambda^i\right)=\sum
      L^j_iY^*_j\lambda^{-1-i}.
      $$
We will need the following lemmas.

\ble\cite{adler}
    We have the following natural identifications:
    $$L\delta _+=L(\delta^*)_- \ \mathrm{and\ } L\delta _-=L(\delta^*)_+.
    $$
\ele

\ble\cite[Lem.~4.1]{sts}\label{lem1}
    Let
    $\varphi$ be an $\mathrm{Ad}$-invariant polynomial on $\delta$. Then
      $$\varphi_{m,n}[L(\lambda)]=
         \mathrm{Res}_{\lambda=0}(\lambda^{-n}\varphi(\lambda^mL(\lambda)))
      $$
    is an $\mathrm{Ad}$-invariant polynomial on $L\delta $ for $m, n\in\mathbb
    Z.$
\ele

As a double Lie algebra,
$\delta$  has an   Ad-invariant polynomial,
the quadratic polynomial
       $$\psi(Y)=\langle Y,Y\rangle       $$
associated to the natural pairing.
   Let $Y_{l+i}=Y^*_i$ for $i\in\{1,\ldots ,l =\mathrm{dim}({\mathfrak g})
    \}$, so elements of $L\delta$ can be written
 $L(\lambda)=\sum\limits_{j=1}^{2l} \sum\limits_{i=-M}^N
    L_i^jY_j\lambda^i$.
Then the Ad-invariant polynomials
\begin{equation}\label{psimn1}
       \psi_{m,n}(L(\lambda))=\mathrm{Res}_{\lambda=0}(\lambda^{-n}
       \psi(\lambda^mL(\lambda))),
\end{equation}
defined as in Lemma \ref{lem1} are given by
\begin{equation}\label{psimn2}
       \psi_{m,n}(L(\lambda))=2\sum
       \limits_{j=1}^l\sum\limits_{i+k-n+2m=-1} L_i^jL_k^{j+l}.
\end{equation}
Note that
    powers of $\psi$ are also $\mathrm{Ad}$-invariant polynomials on
    $\delta $, so
\begin{equation}\label{psimnk}
      \psi^k_{m,n}(L(\lambda))=\mathrm{Res}_{\lambda=0}(\lambda^{-n}\psi^k(\lambda^mL(\lambda)))
\end{equation}
    are  $\mathrm{Ad}$-invariant polynomials on $L\delta $.
It would be interesting to classify all Ad-invariant polynomials on $L\delta$
in general.

\section{The Lax pair equation}\label{s:7}
Let
  $P_+$, $P_-$ be endomorphisms of a Lie algebra $\mathfrak h$ and set $R =
  P_+-P_-.$  Assume
that
     $$[X,Y]_R=[P_+X,P_+Y]-[P_-X,P_-Y]
     $$
is a Lie bracket on $\mathfrak h$.
From \cite[Theorem 2.1]{sts}, the equations of motion induced by a
Casimir (i.e. Ad-invariant) function $\varphi$ on  ${\mathfrak h}^*$
are given by
\begin{equation}\label{e:5.1}
    \frac{dL}{dt}=-\mathrm{ad}^*_{\mathfrak h}M\cdot L,
\end{equation}
for $L\in\mathfrak h^*,$
where
$ M=\frac{1}{2}R(d\varphi(L))\in\mathfrak h.$

Now we take $\mathfrak h=(L\delta)^*=L(\delta^*)$, with $\delta$ a
finite dimensional Lie algebra and with the understanding that
$(L\delta)^*$ is the graded dual with respect to the standard
$\Z$-grading on $L\delta.$ Let $P_\pm$
 be the projections of $L\delta^*$ onto $L\delta^*_\pm$.
After identifying $L\delta ^*=L\delta $
and $\mathrm{ad}^*=-\mathrm{ad}$ via the map $I$  in (\ref{I}),
the equations of motion (\ref{e:5.1})
 can be written in Lax pair form
\begin{equation}\label{e:5.2}
    \frac{{d}L}{{d}t}=[M,L],
\end{equation}
where $ M=\frac{1}{2}R(I(d\varphi(L(\lambda))))\in L\delta,$ and
$\varphi$ is a Casimir function on $L\delta^* = L\delta$
\cite[Theorem 2.1]{sts}. Finding a solution for (\ref{e:5.2})
reduces to the Riemann-Hilbert (or Birkhoff) factorization problem.
The following theorem is a corollary of \cite[Theorem 4.37]{adler}
  \cite[Theorem 2.2]{sts}.

\begin{theorem}\label{t:7.4}
   Let $\varphi$ be a Casimir function on $L\delta$ and set
   $X=I(d\varphi(L(\lambda)))\in L\delta$,
   for
   $L(\lambda) = L(0)(\lambda)\in L\delta$.
   Let
   $g_{\pm}(t)$ be the smooth curves in $L\tilde G$
   which solve the factorization
   problem
   $$\exp(-tX)=g_-(t)^{-1}g_+(t),$$
   with $g_{\pm}(0)=e$, and with $g_+(t) = g_+(t)(\lambda)$
   holomorphic in $\lambda\in {\mathbb C}$ and
   $g_-(t)$ a polynomial in $1/\lambda$ with no constant term.
   Let
   $M=\frac{1}{2}R(I(d\varphi(L(\lambda))))\in L\delta$.
   Then the integral curve $L(t)$ of the Lax pair equation
      $$\frac{d L}{d t}=[L, M]
      $$
   is given by
\begin{equation}\label{quick}
   L(t)=\mathrm{Ad}_{L\tilde G}g_{\pm}(t)\cdot L(0).
      \end{equation}
\end{theorem}

\vspace*{12pt}


This Lax pair equation projects to a Lax pair equation on the loop algebra of
  the original Lie
  algebra $\mathfrak g.$
  Let $\pi_1$ be either the projection of  $\tilde G$ onto $G$ or its
  differential from
$\delta$ onto $\mathfrak g$.
  This extends to a projection of $L\delta$ onto
 $L\mathfrak g$.
  The projection of (\ref{e:5.2}) onto $L\mathfrak g$ is
\begin{equation}\label{e:pi1}
  \frac{d(\pi_1( L(t)))}{dt}=[\pi_1(L),\pi_1(M)],
\end{equation}
since $\pi_1 = d\pi_1 $ commutes with the bracket. Thus the
equations of motion (\ref{e:5.2}) induce a Lax pair equation on
$L\mathfrak g$, although this is not the equations of motion for a
Casimir on $L \mathfrak g.$

\begin{theorem}\label{t:4.2} The Lax pair equation of Theorem \ref{t:7.4} projects to a Lax
  pair equation on $L\mathfrak g.$
\end{theorem}

\begin{remark}
  The content of this theorem is that a Lax pair equation on the Lie algebra
  of a semi-direct product $G\ltimes G'$ evolves on an adjoint orbit, and the
  projection onto $\mathfrak g$ evolves on an adjoint orbit and is still in
  Lax pair form.  Lax pair equations often appear as equations of motion for
  some Hamiltonian, but the projection may not be the equations of motion for
  any function on the smaller Lie algebra.  We thank B. Khesin for this
  observation.
\end{remark}

\vspace*{12pt}
When $\psi_{m,n}$ is the Casimir function on
$L\delta$ given by
    (\ref{psimn1}), $X$ can be written nicely in terms of $L(\lambda)$.
\begin{proposition}\label{p:5.3}
    Let $X=I(d \psi_{m,n}(L(\lambda)))$.
    Then
\begin{equation}\label{e:5.3}
    X=2\lambda^{-n+2m} L(\lambda).
\end{equation}

\end{proposition}

\begin{proof}
Write $
      L(\lambda) =\sum\limits_{i,j}L_i^j\lambda^{i}Y_j.$
  By formula (\ref{psimn2}), we have
  \begin{equation}\label{e:partialpsimn}
    \frac{\partial \psi_{m,n}}{\partial L_p^t}= \left\{
    \begin{array}{cc}
    2L_{n-1-2m-p}^{t+l}, \ \ \ \text{if} \ \ t\leq l\\
    2L_{n-1-2m-p}^{t-l}, \ \ \ \text{if} \ \ t>l.\\
    \end{array}
    \right.
  \end{equation}
  Therefore
      \begin{eqnarray}
           \nonumber  X&=&I(d\psi_{m,n}(L(\lambda)))
            =\sum\limits_{p,t}\frac{\partial \psi_{m,n}}{\partial L_p^t}\lambda^{-1-p}Y^*_t\\
          \nonumber   &=&
            2\lambda^{-n+2m}\sum\limits_p\left(
\sum\limits_{t=1}^{l}L^{t+l}_{n-1-2m-p}Y_{t+l}\lambda^{n-1-2m-p}
            +\sum\limits_{t=l+1}^{2l}
      L^{t-l}_{n-1-2m-p}Y_{t-l}\lambda^{n-1-2m-p}\right)\\
           \nonumber  &=&2 \lambda^{-n+2m}L(\lambda).
       \end{eqnarray}
\end{proof}

\section{The main theorem for Hopf algebras}\label{hopf}

In this section we give formulas for the Birkhoff decomposition
of  a loop in the Lie group of characters of a Hopf algebra
and produce the Lax pair equations associated to the
Birkhoff decomposition. 
We present two approaches, both motivated by the Connes-Kreimer Hopf algebra
of 1PI Feynman graphs.  First, in analogy to truncating Feynman integral
calculations at a certain loop level, we truncate a (possibly infinitely generated)
Hopf algebra to a finitely generated Hopf algebra, and solve Lax pair
equations on the finite dimensional piece (Theorem~\ref{t:8.2fg}).  
We also discuss the compatibility
of solutions related to different truncations.  Second, we solve a Lax pair
equation associated to the full Hopf algebra, but for a restricted family
of Casimirs (Theorem~\ref{t:8.2}).

 Let ${\mathcal H}=({\mathcal H}, 1, \mu,\De,\ep,S)$ be a graded
connected Hopf algebra over ${\mathbb C}$. Let $\mathcal A$ be a
unital commutative algebra with  unit $1_\mathcal A$.  Unless stated otherwise,
$\mathcal{A}$ will be
 the algebra of Laurent series; the only other occurrence in this paper is
 ${\mathcal A} = \CC.$ 

\begin{defn}
  The {\bf character group} $G_\mathcal A$ of the  Hopf algebra ${\mathcal H}$ is
  the set of algebra morphisms $\phi:{\mathcal H}\to\mathcal{A}$
  with $\phi(1)=1_\mathcal{A}.$
  The group law is given by the convolution product
    $$
      (\psi_1\star\psi_2)(h)=\langle  \psi_1\otimes\psi_2,\De h\rangle;$$
  the unit element
  is $\ep$.
\end{defn}

\begin{defn}
  An $\mathcal A$-valued {\bf infinitesimal character} of a Hopf algebra ${\mathcal H}$ is a
  ${\mathbb C}$-linear map $Z:{\mathcal H}\to\mathcal{A}$ satisfying
  $$\langle Z,hk\rangle =\langle Z, h\rangle \varepsilon(k)+\varepsilon(h)
  \langle Z,k\rangle.$$ The set of infinitesimal characters is denoted
  by $\mathfrak{g}_\mathcal{A}$ and is endowed with a Lie algebra bracket:
  $$[Z,Z']=Z\star Z'-Z'\star Z,\ \  \mathrm{for\ }Z,\ Z'\in\mathfrak{g}_\mathcal{A}, $$
  where
  $\langle Z\star Z',h\rangle=\langle Z\otimes Z',\Delta(h)\rangle$.
  Notice that $Z(1)=0$.
\end{defn}

For a finitely generated Hopf algebra, $G_\CC$ is a Lie group with Lie algebra
${\mathfrak g}_\CC$, and for any Hopf algebra and any ${\mathcal A}$, the same
is true at least formally.

We recall that $\delta=\mathfrak{g}_{\mathcal{{\mathbb
C}}}\oplus\mathfrak{g}_{\mathcal{{\mathbb C}}}^{*}$ is the double of
$\mathfrak{g}_{\mathbb C}$ and the $\mathfrak{g}_{\mathbb C}^{*}$ is
the graded dual of $\mathfrak{g}_{\mathbb C}$. We consider the
algebra
  $\Omega\delta = \delta\otimes\mathcal{A}$
of formal Laurent series with values in
 $\delta$
    $$\Omega\delta=\{L(\lambda)=\sum\limits_{j=-N}^\infty \lambda^jL_j \ |
                    \ L_j\in\delta, N\in \Z\}.
    $$
The natural Lie bracket on $\Omega\delta $ is
     $$\left[\sum \lambda^iL_i,\sum \lambda^j L_j'\right]=
        \sum\limits_k \lambda^k\sum\limits_{i+j=k}[L_i,L_j'].
     $$
Set
\begin{eqnarray*}\Omega\delta _+ &=&
    \{L(\lambda)= \sum\limits_{j=0}^\infty
       \lambda^jL_j \ |  \ L_j\in\delta\}\\
     \Omega\delta _- &=& \{L(\lambda)=\sum\limits_{j=-N}^{-1}
      \lambda^jL_j \ |  \    L_j\in\delta, N\in \Z^+\}.
\end{eqnarray*}

Recall that for any Lie group $K$, a loop  $L(\lambda)$ with values
in $K$ has a Birkhoff decomposition if $L(\lambda) = L(\lambda)
_-^{-1}L(\lambda)_+$ with $L(\lambda)_-^{-1}$ holomorphic in
$\lambda^{-1} \in \mathbb P^1 -\{0\}$ and $L(\lambda)_+$ holomorphic
in $\lambda \in\mathbb P^1 - \{\infty\}.$  In the next lemma,
$\tilde G$ refers  to $G\ltimes_\theta \mathfrak g^*$ as in
Prop.~\ref{prop22}.

We prove the existence of a Birkhoff
decomposition for any element
 $(g,\alpha)\in \Omega\tilde G$.

\begin{theorem}\label{t:omegatildeg}
   Every
   $(g,\alpha)\in\Omega\tilde
    G=G_\mathcal{A}\ltimes_{Ad^*_{G_{\mathcal A}}}
    \mathfrak{g}_{\mathcal{A}}^{*}
   $
    has a Birkhoff decomposition
  $(g,\alpha)=(g_-,\alpha_-)^{-1}(g_+,\alpha_+)$ with
  $(g_+,\alpha_+)$ holomorphic in $\lambda$ and
  $(g_-,\alpha_-)$  a polynomial in $\lambda^{-1}$
  without constant term.
\end{theorem}
\begin{proof}
   We recall that
  $(g_1,\alpha_1)(g_2,\alpha_2)=
  (g_1g_2,\alpha_1+\mathrm{Ad}^*(g_1)(\alpha_2)).$
Thus\\
  $
    (g,\alpha)=(g_-,\alpha_-)^{-1}(g_+,\alpha_+)
      \text{ if and only if }
      g=g_-^{-1}g_+
    \text{ and }
    \alpha=\mathrm{Ad}^*(g_-^{-1})(-\alpha_-+\alpha_+).
$
    Let
       $g=g_-^{-1}g_+$
    be the Birkhoff decomposition of $g$ in $ G_\mathcal{A}$
    given in \cite{ck1,egk,man}.
    Set $\alpha_+=P_+(\mathrm{Ad}^*(g_-)(\alpha))$
    and
       $\alpha_-=-P_-(\mathrm{Ad}^*(g_-)(\alpha))$, where $P_+$ and
       $P_-$ are the holomorphic and pole part,
      respectively.
    Then for this choice of $\alpha_+$ and $\alpha_-$, we have
    $(g,\alpha)=(g_-,\alpha_-)^{-1}(g_+,\alpha_+)$. Note that the
    Birkhoff decomposition is unique.
\end{proof}

For a finitely generated Hopf algebra, we can apply Theorems~\ref{t:7.4},
  \ref{t:4.2} to produce a Lax pair equation on $L\delta$ and on the loop
  space of infinitesimal characters $L\mathfrak{g}$.  However, the common Hopf
  algebras of 1PI Feynman diagrams and rooted trees are not finitely generated.

As we now explain, we can truncate the Hopf algebra to a finitely generated Hopf algebra, and use
the Birkhoff decomposition to solve a Lax pair equation on the infinitesimal
character group of the truncation.  A graded Hopf algebra
$\mathcal H=\oplus_{n\in\mathbb N} \mathcal H_n$ is said to be of {\bf finite type} if each
homogeneous component $\mathcal H_n$ is a finite dimensional vector space.
Let $\mathcal B=\{T_i\}_{i\in\mathbb N}$ be a minimal set of
homogeneous generators of the Hopf algebra $H$ such that
$\deg(T_i)\leq\deg(T_j)$ if $i<j$ and such that $T_0=1$.
For $i>0$, we define the
$\mathbb{C}$-valued infinitesimal character $Z_i$ on generators by
$Z_i(T_j)=\delta_{ij}.$
The Lie algebra of infinitesimal characters
$\mathfrak g$ is a graded Lie algebra  generated by
$\{Z_i\}_{i>0}$.
Let $\mathfrak{g}^{(k)}$
 be the vector space generated by $\{Z_i\ | \ \deg(T_i)\leq k\}$. We
define $\deg(Z_i)=\deg(T_i)$ and set
$$[Z_i,Z_j]_{\mathfrak{g}^{(k)}}=\left\{
\begin{array}{cc}
 [Z_i,Z_j] & \text{if }\deg(Z_i)+\deg(Z_j)\leq k \\
 0 & \text{if }\deg(Z_i)+\deg(Z_j)>k 
\end{array}\right.$$
We identify $\varphi\in G_{\mathbb C}$ with
$\{\varphi(T_i)\}\in\mathbb C^{\mathbb N}$ and on $\mathbb
C^{\mathbb N}$ we set a group law given by
$\{\varphi_1(T_i)\}\oplus\{\varphi_2(T_i)\}=\{(\varphi_1\star\varphi_2)(T_i)\}$.
 $G^{(k)}=\{ \{\varphi(T_i)\}_{\{i\, |\, \deg(T_i)\leq
k\}}\ | \ \varphi\in G_{\mathbb C}\}$ is a finite dimensional Lie
subgroup of $G_{\mathbb C}=(\mathbb C^{\mathbb N},\oplus)$ and the Lie algebra
of $G^{(k)}$ is $\mathfrak{g}^{(k)}$.  There is no loss of information under this
identification, 
 as $\varphi(T_iT_j)=\varphi(T_i)\varphi(T_j)$.

Let $\delta^{(k)}$ be the double Lie algebra of $\mathfrak{g}^{(k)}$
and let $\tilde G^{(k)}$ be the simply connected Lie group with
$\mathrm{Lie}(\tilde G^{(k)})=\delta^{(k)}$   as in Proposition
\ref{prop22}. The following theorem is a restatement of Theorem
\ref{t:7.4} in our new stage.

\begin{theorem}\label{t:8.2fg}
   Let $\mathcal H=\oplus_{n}\mathcal 
H_n$ be a graded connected Hopf algebra of finite type, and let
     $\psi:L\delta^{(k)}\to{\mathbb C}$ be a Casimir function
     (e.g. $\psi(L ) = \psi_{m,n}(L(\lambda))=\mathrm{Res}_{\lambda=0}
       (\lambda^m\psi(\lambda^n L(\lambda)))$ with
     $\psi:\delta^{(k)}\times\delta^{(k)}\to{\mathbb C}$   the natural paring of
$\delta^{(k)}$).
 Set $
  X=I(d\psi(L_0 ))$ for $L_0 \in L\delta^{(k)}$.
  Then the solution in $L\delta^{(k)}$ of
 \begin{equation}\label{8:1fg}
        \frac{dL}{dt}=[L,M]_{L\delta^{(k)}}, \ \ \ M=\frac{1}{2}
        R(I(d\psi(L)))
   \end{equation}
   with initial condition $L(0)=L_0$ is given by
   \begin{equation}\label{e:8.3fg}
        L(t)=\mathrm{Ad}_{L\tilde G^{(k)}}g_{\pm}(t)\cdot L_0,
   \end{equation}
   where $\exp(-tX)$ has the Connes-Kreimer Birkhoff factorization\\
        $\exp(-tX)=g_-(t)^{-1}g_+(t)$.
\end{theorem}

\begin{remark}
(i) If $L_0\in L\delta$,  there exists $k\in\mathbb N$ such that
$L_0\in L\delta^{(k)}$. Indeed $L_0\in L\delta$ is generated over
$\mathbb C[\lambda,\lambda^{-1}]$ by a finite number of
 $\{Z_i\}$, and we can choose
  $k\geq\max\{\deg(Z_i)\}$.

(ii) 
While the Hopf algebra of rooted trees and the Connes-Kreimer Hopf algebra
of 1PI Feynman diagrams satisfy the hypothesis of Theorem
\ref{t:8.2fg}, the Feynman rules
character does not lie in $L\tilde G$, as explained below.

\end{remark}

In the next sections, we will investigate the relationship between
the Lax pair flow $L(t)$ and the Renormalization Group Equation. In
preparation, we project from $L\delta^{(k)}$ to
$L\mathfrak{g}^{(k)}$ via $\pi_1$ as in \S4.

\begin{corollary}\label{tc:8.2fg}
Let $\psi $ be a Casimir function on $L\delta^{(k)}$.
 Set $L_0\in L\mathfrak{g}^{(k)}\subset L\delta^{(k)}$,
 $X=\pi_1(I(d\psi(L_0)))$.
 Then the solution of the following
equation in $L\mathfrak{g}^{(k)}$
\begin{equation}\label{c:e:8.1}
        \frac{dL}{dt}=[L,M_1]_{L\mathfrak{g}^{(k)}}, \ \ \ M_1=\pi_1(\frac{1}{2}
        R(I(d\psi (L ))))
   \end{equation}
   with initial condition $L(0)=L_0$ is given by
   \begin{equation}\label{c:e:8.3}
        L(t)=\mathrm{Ad}_{L G^{(k)}}g_{\pm}(t)\cdot L_0,
   \end{equation}
   where $\exp(-tX)$ has the Connes-Kreimer Birkhoff factorization
   in $L\mathfrak{g}^{(k)}$
        $$\exp(-t X)=g_-(t)^{-1}g_+(t).$$
\end{corollary}

\begin{remark}\label{lastrem}
(i)  
For Feynman graphs, this truncation corresponds to halting
calculations after a certain loop level.  From our point of view, this
truncation is somewhat crude.  $\mathfrak g^{(k)}$ is not a subalgebra of
$\mathfrak g$, and if $k < \ell$, $\mathfrak g^{(k)}$ is not a subalgebra of
$\mathfrak g^{(\ell)}$.  Although the Casimirs $\psi_{m,n}$ and the exponential map restrict
well from $\mathfrak g$ to
$\mathfrak g^{(k)}$, the Birkhoff decomposition $\exp(-tX)$ of $X\in
L{\mathfrak g}^{(k)}$ is very different from the Birkhoff decompositions in
$L\mathfrak g, L\mathfrak g^{(\ell)}$.  In fact, if $g\in G^{(k)}$ has
Birkhoff decomposition $g = g_-^{-1}g_+$ in $G$, there does not seem to be
$f(k)\in \mathbb N$ such that $g_\pm\in G^{(f(k))}.$
Nevertheless, in the last section we will follow standard procedure and
present calculations of truncated Hopf algebras.

(ii)It would interesting to know, especially for the 
 Hopf algebras of Feynman graphs or rooted trees,
 whether there exists a larger connected graded Hopf algebra ${\mathcal H}'$
 containing ${\mathcal H}$ such that the associated infinitesimal Lie algebra
 $\mathrm{Lie}(G'_\mathbb{C})$ is the double $\delta$. This would provide a Lax pair equation
 associated to an equation of motion on the infinitesimal Lie algebra
 of ${\mathcal H}'$. 
The most natural candidate, the Drinfeld double $\mathcal D({\mathcal H})$ of ${\mathcal H}$,
 does not work since the dimension of the Lie algebra associated to
 $\mathcal D({\mathcal H})$ is larger than the dimension of $\delta$.
\end{remark}

In \cite{ck1}, Connes and Kreimer give a Birkhoff decomposition for
the character group of the  Hopf algebra of 1PI graphs, and in
particular for the Feynman rules character $\varphi(\lambda)$ given
by minimal subtraction and dimensional regularization. 
The truncation process treated above does not handle the Feynman rules
character, as
the Feynman
rules character  and the toy model
character of the Hopf algebra of rooted trees considered in \S8 are
not  polynomials in $\lambda,\lambda^{-1}$, but Laurent series in
$\lambda$. Thus Corollary \ref{tc:8.2fg}
does not apply, as in our notation $\log(\varphi(\lambda)) \in
\Omega\mathfrak{g} \setminus L\mathfrak{g}$.
This and Remark \ref{lastrem}(i) force us to consider a direct approach in
$\Omega\mathfrak{g}$ as in next theorem.  However, we cannot expect that
the Lax pair equation is associated to  any Hamiltonian
equation, and we replace Casimirs with Ad-covariant functions.

\begin{defn} \cite{suris}
Let $G$ be a Lie group with Lie algebra $\mathfrak{g}$.
 A map $f:\mathfrak{g}\to \mathfrak{g}$ is  $\mathrm{Ad}$-{\it covariant} if
$\mathrm{Ad}(g)(f(L))=f(\mathrm{Ad}(g)(L))$
for all $g\in G$, $L\in\mathfrak{g}$.
\end{defn}

\begin{theorem}\label{t:8.2}
 Let ${\mathcal H}$ be a connected graded commutative Hopf algebra with
 $\mathfrak{g}_{\mathcal A}$ the associated Lie algebra of infinitesimal
 characters with
 values in Laurent series.
 Let $f:\mathfrak{g}_{\mathcal A}\to\mathfrak{g}_{\mathcal A}$
 be an $\mathrm{Ad}$-covariant
 map.
 Let $L_0\in\mathfrak{g}_{\mathcal A}$ satisfy
 $[f(L_0),L_0]=0$.
 Set $X=f(L_0)$.
 Then the solution of
 \begin{equation}\label{8:1}
        \frac{dL}{dt}=[L,M], \ \ \ M=\frac{1}{2}
        R(f(L))
   \end{equation}
   with initial condition $L(0)=L_0$ is given by
   \begin{equation}\label{e:8.3}
        L(t)=\mathrm{Ad}_{G}g_{\pm}(t)\cdot L_0,
   \end{equation}
   where $\exp(-tX)$ has the Connes-Kreimer Birkhoff factorization\\
        $\exp(-tX)=g_-(t)^{-1}g_+(t)$.
\end{theorem}

\begin{proof}
The proof is similar to    \cite[Theorem 2.2]{sts}.
First notice that
\begin{eqnarray*}
  \frac{d}{dt}\left({\Ad}(g_-(t)^{-1}g_+(t))\cdot
  L_0\right) &=&
  \frac{d}{dt}(\exp(-tX)L_0\exp(tX))\\
  &=& -\exp(-tX)XL_0\exp(tX)+\exp(-tX)L_0X\exp(tX)\\
  &=& \exp(-tX)[X,L_0]\exp(tX)=0,
\end{eqnarray*}
which implies ${\Ad}(g_-(t)^{-1}g_+(t))\cdot
  L_0=L_0$ and  ${\Ad}(g_-(t) )\cdot  L_0={\Ad}(g_+(t))\cdot  L_0$.
Set $L(t)={\Ad}(g_\pm(t) )\cdot  L_0 = g_\pm(t)L_0g_\pm(t)^{-1}$.
As usual,
$$
 \frac{dL}{dt}
 = \left[\frac{d g_\pm(t)}{dt}g_\pm(t)^{-1},L(t)\right],
$$
so
$$
 \frac{dL}{dt}=\frac{1}{2}\left[\frac{d g_+(t)}{dt}g_+(t)^{-1}+\frac{d
 g_-(t)}{dt}g_-(t)^{-1},L(t)\right]. 
$$
The Birkhoff factorization 
 $g_+(t)=g_-(t)\exp(-tX)$ gives
 $$\frac{dg_+(t)}{dt}
    =\frac{dg_-(t)}{dt}\exp(-tX)+g_-(t)(-X)\exp(-tX),$$
and so
$$\frac{dg_+(t)}{dt}g_+(t)^{-1}
    =\frac{dg_-(t)}{dt}g_-(t)^{-1}+g_-(t)(-X)g_-(t)^{-1}.$$
Thus
\begin{eqnarray*} 
2M &=& R(f(L(t)))=R(f({\Ad}( g_-(t))\cdot L_0))=R({\Ad}( g_-(t))\cdot f(L_0))\\
&=& R({\Ad}( g_-(t))\cdot X)) = 
-R(\frac{dg_+(t)}{dt}g_+(t)^{-1})
    +R(\frac{dg_-(t)}{dt}g_-(t)^{-1})\\
&= &   -\frac{dg_+(t)}{dt}g_+(t)^{-1}
    -\frac{dg_-(t)}{dt}g_-(t)^{-1}.
\end{eqnarray*}
Here we use $(\frac{dg_\pm(t)}{dt}g_\pm(t)^{-1})(x)\in\mathcal
A_\pm$ for
$x\in {\mathcal H}$.
 Thus $
 \frac{dL}{dt}= [L,M].$
\end{proof}

If $f:\mathfrak{g}_\mathcal{A}\to\mathfrak{g}_\mathcal{A}$ is 
given by $f(L)=2\lambda^{-n+2m}L$,
 then $f$ is 
$\mathrm{Ad}$-covariant and 
$[f(L_0),L_0]=[2\lambda^{-n+2m}L_0,L_0]=0$.  

\begin{corollary}\label{tc:8.2}
Let ${\mathcal H}$ be a connected graded commutative Hopf algebra with 
 $\mathfrak{g}_{\mathcal A}$  the Lie algebra of infinitesimal
characters with
values in Laurent series.
Pick $L_0\in\mathfrak{g}_{\mathcal A}$ and set
$X=2\lambda^{-n+2m}L_0$. Then the solution of
 \begin{equation}\label{8:1gen}
        \frac{dL}{dt}=[L,M], \ \ \ M=
        R(\lambda^{-n+2m} L)
   \end{equation}
   with initial condition $L(0)=L_0$ is given by
   \begin{equation}\label{e:8.3gen}
        L(t)=\mathrm{Ad}_{ G_\mathcal{A}}g_{\pm}(t)\cdot L_0,
   \end{equation}
   where $\exp(-tX)$ has the Connes-Kreimer Birkhoff factorization\\
        $\exp(-tX)=g_-(t)^{-1}g_+(t)$.
 \end{corollary}

\begin{remark}\label{FRrmk}
     Let $\varphi$ be the Feynman rules character. We can find the Birkhoff factorization of $\varphi$ itself
       within this framework by adjusting the initial condition.
     Namely, set
     $L_0(\lambda)=\frac{1}{2}\lambda^{n-2m}\exp^{-1}(\varphi(\lambda)).$
     Then $\exp(X) =  \varphi$ by Prop.~\ref{p:5.3}, so the solution
     of \eqref{8:1gen} involves the Birkhoff factorization $ \varphi =
     g_-(-1)^{-1}g_+(-1)$.
     Namely, we have
     $$L (-1 )= \frac{\lambda^{n-2m}}{2}
     \mathrm{Ad}_{ G_\mathcal{A}}g_\pm (-1 )
     \exp^{-1}( \varphi).$$
\end{remark}

\section{The Connes-Kreimer $\beta$-function}

The flow of characters usually considered in quantum field theory is the
renormalization group flow (RGF).  In contrast, the Lax pair flow lives on the Lie
algebra of the character group.  Since the $\beta$-function of the RGF is an
element of the Lie algebra of the $\CC$-valued characters, it is natural to
examine the relationship 
 between the Lax pair equations and
the $\beta$-function. In this section, we continue to work in the general
setup of Hopf algebras and character groups.

Here we consider two flows for the
$\beta$-function. First, we extend the (scalar) beta function of a local character
$\varphi$ (see (\ref{locality})) to 
an infinitesimal  character $\tilde \beta_\varphi$ (Lemma \ref{l:holo}).   
This ``beta character'' 
has already appeared in the literature:
$\tilde\beta_\varphi = \lambda\tilde R(\varphi)$, in the language of \cite{man}
explained below (Lemma \ref{lemma:6.6}), but it seems worth highlighting. 
For certain Casimirs, we show that the
beta character is a fixed point of the Lax pair flow (Theorem
\ref{t:4.5.3}).

It is more important and more difficult  to consider the flow of the 
$\beta$-function itself. Namely, given a character $\varphi$, we can set $L_0 =
\log(\varphi)$ and study the $\beta$-functions of the  characters $\varphi(s)
= \exp(L(s)).$
 In Theorem \ref{t:flow}, we give a
differential equation for $\beta_{\varphi(s)}.$

To define the beta character, we recall material from
 \cite{ck2,ef-manchon,man}.
Throughout this
section, $\mathcal A$  denotes the algebra of Laurent series.

Let $\mathcal H=\bigoplus\limits_{n}\mathcal H_n$ be a connected
graded Hopf algebra. Let $Y$ be the  biderivation on $\mathcal H$ given
on homogeneous elements by
$$Y:\mathcal H_n\to \mathcal H_n,\ \ \ \ \ Y(x)=nx\ \ \ \text{for }
x \in\mathcal H_n.$$

\begin{defn} \label{def:6.1} \cite{man}
 We define the bijection $\tilde R:G_{\mathcal A}\to\mathfrak g_{\mathcal A}$ by
 $$\tilde R(\varphi)=\varphi^{-1}\star(\varphi\circ Y).$$
\end{defn}
\noindent
 Consider the semidirect
 product Lie algebra
 $\tilde{\mathfrak g}_{\mathcal A}
 =\mathfrak g_{\mathcal A}\rtimes{\mathbb C}\cdot Z_0,$ where $Z_0$ acts via
$[Z_0,X]=X\circ Y$ for  $X\in\mathfrak g_{\mathcal A}$.
 Let $\{\theta_t\}_{t\in\mathbb C}$ be the one-parameter group of
 automorphisms of
$\mathcal
 H$ given by
 $$\theta_t(x)=e^{nt}x, \text{   for }x\in\mathcal H_n.$$
 Then  $\varphi\ \ 
\mapsto
\varphi\circ\theta_t$ is an automorphism of
 $G_{\mathcal A}$.
 Let $\tilde{G}_{\mathcal A}$ be the semidirect product
 $$\tilde{G}_{\mathcal A}=G_{\mathcal A}\rtimes\mathbb C ,$$
 with the action of  $\mathbb C $ on $G_{\mathcal A}$ given by
 $\varphi\cdot t=\varphi\circ\theta_t$.
 $G_{\mathcal A}$ has Lie algebra $\mathfrak g_{\mathcal A}$.

We  now define a second action of $\mathbb C$ on
 $G_{\mathcal A}$. For $t\in\mathbb C$ and $\varphi\in G_{\mathcal A}$ we
 define $\varphi^t(x)$ on an homogeneous element $x\in \mathcal H$ by
 $$\varphi^t(x)(\lambda)=e^{t\lambda |x|} \varphi(x)(\lambda),$$
 for any $\lambda\in\mathbb C$, where $|x|$ is the degree of
 $x$.

\begin{defn} 
 Let
\begin{equation}\label{locality}
G^{\Phi}_{\mathcal A}=\{\varphi\in G_{\mathcal A}\ \big|\ \frac{ \ \ d}{dt}
(\varphi^t)_-=0\},
\end{equation}
  be the set
of characters with the negative part of the Birkhoff
  decomposition independent of $t$. Elements of  $ G^{\Phi}_{\mathcal A}$
are called  {\it local characters}.
\end{defn}

  The dimensional regularized
  Feynman rule character $\varphi $ is local.
  Referring to \cite{ck2,ef-manchon}, the physical meaning of locality is that
  the counterterm $\varphi_-$ does not depend on the mass parameter $\mu$:
  $\frac{\partial\varphi_- }{\partial\mu}=0$.

\begin{proposition}[\cite{ck2,man,ef-manchon}]
  Let $\varphi\in G_{\mathcal A}^{\Phi}$.
   Then the limit
  $$F_\varphi(t)=\lim\limits_{\lambda\to 0}\varphi ^{-1}(\lambda)\star
  \varphi^t (\lambda)$$ exists and  is a 
  one-parameter subgroup in $G_{\mathcal A}\cap G_{\mathbb C}$ of scalar valued
  characters of $\mathcal H$.
\end{proposition}
\noindent
  Notice that $(\varphi ^{-1}(\lambda)\star \varphi^t (\lambda))(\Gamma)\in\mathcal A_+$ as
  $$\varphi ^{-1}(\lambda)\star \varphi^t
   (\lambda)=\varphi_+^{-1}\star\varphi_-\star
   (\varphi^t)_-^{-1}\star(\varphi^t)_+=
    \varphi_+^{-1}\star(\varphi^t)_+.$$
\begin{defn}
  For $\varphi\in G^\Phi_{\mathcal A}$, the $\beta$-function of
  $\varphi$ is defined to be
  $\beta_\varphi=-(\mathrm{Res}(\varphi_-))\circ Y)$.
 \end{defn}
We have \cite{ck2}
  $$\beta_\varphi={d\over dt}\Big|_{t=0} F_{\varphi_-^{-1}}(t),$$
where $F_{\varphi_-^{-1}}$, the one-parameter subgroup associated to
$\varphi_-^{-1}$,  also belongs to $G^\Phi_\mathcal{A}$.

 To relate the
$\beta$-function $\beta_\varphi\in {\mathfrak g}_\CC$ to our Lax pair equations,
 which live on ${\mathfrak g}_{\mathcal A}$, we can either consider
 ${\mathfrak g}_\CC$ as a subset of ${\mathfrak g}_{\mathcal A}$, or we can
 extend $\beta_\varphi$ to an element of ${\mathfrak g}_{\mathcal A}.$  Since
 ${\mathfrak g}_\CC$ is not preserved under the Lax pair flow, we take the
 second approach.

\begin{defn}
  For $\varphi\in G_{\mathcal A}^{\Phi}$, $x\in H$, set
  $$\tilde{\beta}_\varphi(x)(\lambda) ={ d\over dt}\Big|_{t=0}
  (\varphi^{-1}\star\varphi^t)(x)(\lambda).$$
\end{defn}

 The following lemma establishes that $\tilde\beta$ is an infinitesimal
character. 

\begin{lemma}\label{l:holo}
  Let $\varphi\in G_{\mathcal A}^{\Phi}$.

 i) $\tilde{\beta}_\varphi $ is an infinitesimal character in
  $\mathfrak g_{\mathcal A}$.\\

  ii) $\tilde{\beta}_\varphi $ is holomorphic (i.e.
  $\tilde{\beta}_\varphi(x)\in\mathcal A_+ $ for any $x$).
\end{lemma}
\begin{proof}
 i) For two homogeneous elements $x,y\in\mathcal H$, we have:
  $$\varphi^t(xy)=e^{t|xy|\lambda}\varphi(xy)=
  e^{t|x|\lambda}\varphi(x)e^{t|y|\lambda}\varphi(y)=\varphi^t(x)\varphi^t(y).
  $$
 Therefore $\varphi\star\varphi^t\in G_{\mathcal A}$. Since
 $\varphi^{-1}\star\varphi^0=e$ we get
 $$ {d\over dt}\Big|_{t=0} \varphi^{-1}\star\varphi^t\in\mathfrak
 g_{\mathcal A}.$$

ii) Since ${d\over dt}(\varphi^t)_-=0$, we get
  $$\tilde{\beta}_\varphi=(\varphi_+)^{-1}\star\varphi_-\star((\varphi^t)_-)^{-1}\star(\varphi^t)_+
    =(\varphi_+)^{-1}\star(\varphi^t)_+.
  $$
  Then
 $$\tilde{\beta}_\varphi(x)=(\varphi_+)^{-1}(x')(\varphi^t)_+(x'')
 =(\varphi_+)(S(x'))(\varphi^t)_+(x'')
 $$
  Therefore $\tilde{\beta}_\varphi(x)\in\mathcal A_+$.
\end{proof}

\begin{lemma}\label{lemma:6.6} If $\varphi\in G^\Phi_{\mathcal{A}}$ then

(i)
$\tilde\beta_\varphi=\lambda\tilde R(\varphi)$,

(ii)
  $\beta_\varphi=\mathrm{Ad}(\varphi_+(0))(\tilde\beta_\varphi\big|_{\lambda=0})$,

(iii)  $\tilde{\beta}_{\varphi_-}(x)(\lambda = 0) =-\beta_\varphi(x).$
\end{lemma}
\begin{proof} (i)
For $\Delta(x)=x'\otimes x''$, we have
$$\tilde{\beta}_\varphi(x)(\lambda) ={ d\over dt}\Big|_{t=0}
  (\varphi^{-1}\star\varphi^t)(x)(\lambda)=
  \varphi^{-1}(x'){ d\over dt}\Big|_{t=0}(\varphi^t)(x'')$$
  $$=
  \varphi^{-1}(x')\lambda\cdot
  \mathrm{deg}(x'')\varphi(x'')=\lambda\varphi^{-1}(x')\varphi\circ
  Y(x'')=\lambda(\varphi^{-1}\star(\varphi\circ Y))(x)=\lambda\tilde
  R(\varphi)(x)
  .$$

(ii) 
  The cocycle property of $\tilde R$ \cite{ef-manchon}, $\tilde
  R(\phi_1\star\phi_2)=\tilde R(\phi_2)+\phi_2^{-1}\star\tilde
  R(\phi_1)\star\phi_2$, implies that
\begin{eqnarray}\label{e:e:6.6}
  \lambda\tilde R(\varphi)=\lambda\tilde
  R(\varphi_-^{-1}\star\varphi_+)=\lambda\tilde
  R(\varphi_+)+\varphi_+^{-1}\star\lambda\tilde
  R(\varphi_-^{-1})\star\varphi_+.
\end{eqnarray}
  Since $\tilde
  R(\varphi_+)=\varphi_+^{-1}\star(\varphi_+\circ Y)$ is always
  holomorphic and since $\lambda\tilde
  R(\varphi_-^{-1})=\mathrm{Res}(\varphi_-^{-1})\circ
  Y=-\mathrm{Res}(\varphi_-)\circ
  Y=\beta$ by \cite[Theorem IV.4.4]{man},  when we evaluate (\ref{e:e:6.6}) at $\lambda=0$ we get
  $\tilde\beta(\varphi)\big|_{\lambda=0}=\mathrm{Ad}(\varphi_+^{-1}(0))\beta$.

(iii) 
 The Birkhoff decomposition of
 $\varphi_-=(\varphi_-)_-^{-1}\star(\varphi_-)_+$ is given by
 $(\varphi_-)_-=\varphi_-^{-1}$ and $(\varphi_-)_+=\varepsilon$. 
By definition,
 $\beta_{\varphi_-}=-\mathrm{Res}((\varphi_-)_-)\circ Y
 =-\mathrm{Res}(\varphi_-^{-1})\circ Y
 =\mathrm{Res}(\varphi_- )\circ Y=-\beta_\varphi
 $. Applying (ii) to $\varphi_-$, we get
 $$-\beta_{\varphi} = \beta_{\varphi_-}
   =\mathrm{Ad}(\varepsilon\big|_{\lambda=0})
    (\tilde\beta_{\varphi_-}\big|_{\lambda=0})
   =\tilde\beta_{\varphi_-}\big|_{\lambda=0}.
 $$
\end{proof}

  If $\varphi\in G_{\mathcal A}^\Phi$, the Lax pair equation in
  Corollary \ref{tc:8.2} for $L_0=\tilde{\beta}_\varphi $ is
\begin{eqnarray}\label{p:beta-flow-h}
  {d\over ds}\tilde{\beta}_\varphi(s)=[\tilde{\beta}_\varphi(s),M],
\end{eqnarray}
  where $M=R(\lambda^{-n+2m}\tilde{\beta}_\varphi(s))$ and
  the solution is given by
\begin{eqnarray}\label{p:beta-flow}
\tilde{\beta}_\varphi(s)=\mathrm{Ad}(g_+(s))\tilde{\beta}_\varphi(0)
\end{eqnarray}
 for $g_\pm(s)$ given by the Birkhoff decomposition
 $\exp(s\lambda^{-n+2m}\tilde{\beta}_\varphi)=g^{-1}_-(s)\star
  g_+(s).$

The next theorem shows that the $\beta$-function is a fixed point of
the Lax pair flow for certain Casimirs. Of course this is not the
same as having the $\beta$-function a fixed point of the RGF.

 \begin{theorem}\label{t:4.5.3}
  $\tilde{\beta}_\varphi(s)$ and therefore
  $\beta_\varphi(s)=\mathrm{Ad}(\varphi_+(s)\Bigl|_{\lambda=0}
)(\tilde{\beta}_\varphi(s)\Big|_{\lambda=0})$ are constant
  under the Lax flow if $-n+2m\geq 0$.
 \end{theorem}

\begin{proof}  We drop $s$ from the notation. 
  If $-n+2m\geq 0$ then
 $$M=R(  \lambda^{-n+2m}\tilde{\beta}_\varphi)=
\lambda^{-n+2m}\tilde{\beta}_\varphi,$$
 since $\tilde{\beta}_\varphi$ is holomorphic by Lemma \ref{l:holo} and
 Theorem \ref{p:holo}.  (The proof of this Theorem is independent of this section.)
 So the Lax pair equation becomes
 $${d\over ds}\tilde{\beta}_\varphi
 =[ \tilde\beta_\varphi,  \lambda^{-n+2m}\tilde{\beta}_\varphi]
 =  
  \lambda^{-n+2m}[\tilde{\beta}_\varphi, \tilde{\beta}_\varphi]=0.
 $$
\end{proof}

Now we consider the more interesting case of the flow
$\beta_{\psi(s)}$ of the $\beta$-function of exponentiated
infinitesimal characters. We first establish some simple 
properties of $\varphi^t$.

\begin{lemma} Let $\varphi\in G_{\mathcal A}$.

(i)  $(\varphi\star\psi)^t=\varphi^t\star\psi^t$,

(ii) $(\varphi^{-1})^t=(\varphi^t)^{-1}$,
\end{lemma}
\begin{proof}
 We have
 \begin{eqnarray*}
    (\varphi\star\psi)^t(x)&=&e^{t|x|\lambda}(\varphi\star\psi)
    (x)=\sum\limits_{(x)}e^{t|x|\lambda}\varphi(x')\psi(x'') \\
    &=&\sum\limits_{(x)}e^{t(|x'|+|x''|)\lambda}\varphi(x')\psi(x'')
    =e^{t|x'|\lambda}\varphi(x')\ e^{t|x''|\lambda}\psi(x'')
    =\varphi^t(x')\psi^t(x'')\\
     &=&(\varphi^t\star\psi^t)(x).
 \end{eqnarray*}
 Therefore
   $$ \varphi^t\star(\varphi^{-1})^t
    =(\varphi\star\varphi^{-1})^t=\varepsilon^t=\varepsilon=\varphi^t\star(\varphi^t)^{-1},
 $$
so $(\varphi^{-1})^t=(\varphi^t)^{-1}$.
\end{proof}

The exponential map $\exp:{\mathfrak g}_{\mathcal A}\to G_{\mathcal A}$
is a bijection. Therefore, we can transfer the Lax pair flow on ${\mathfrak g}_{\mathcal
  A}$ to a flow on $G_{\mathcal A}$, and study the associated flow of beta characters.


\begin{theorem}\label{t:flow}
 Let $\varphi\in G^\Phi_\mathcal A$. Let 
 $$\dot\psi(s)=[\psi(s),M]$$
 be the Lax pair from Theorem~\ref{t:8.2}
with  $\psi(0)=\psi = \log(\varphi)$.
 Let $\varphi(s)=\exp(\psi(s))$.
 For $$\tilde{\beta}_{\varphi(s)}=
 {d\over dt}\Big|_{t=0}\varphi(s)^{-1}\star(\varphi(s))^t,$$
 we have
\begin{eqnarray}\label{e:t:phi}
 {d\over ds}
 \tilde{\beta}_{\varphi(s)}&=&
 [\tilde{\beta}_{\varphi(s)},\varphi^{-1}(s)\star
 d\exp[\log\varphi(s),M]]\\
&&\qquad  +\lambda(\varphi^{-1}(s)\star d\exp[\log\varphi(s),M])\circ Y.\nonumber
\end{eqnarray}
\end{theorem}
\begin{proof}
Omitting some stars, we have
\begin{eqnarray}\label{e:beta1}
 {d\over ds}
 \tilde{\beta}_{\varphi(s)}(x)
 \nonumber &=&
 {d\over ds} {d\over dt}\Big|_{t=0}(\varphi^{-1}(s)\star\varphi^t(s))(x)\\
 \nonumber &=&
  {d\over dt}\Big|_{t=0}( -\varphi^{-1}(s) {d\over ds}\exp\psi(s)\varphi^{-1}(s)\varphi^t(s)
  +\varphi^{-1}(s) ({d\over ds}\exp\psi(s))^t)(x)\\
 \nonumber&=&
  {d\over dt}\Big|_{t=0}( -\varphi^{-1}(s)
  d\exp\dot\psi(s)\varphi^{-1}(s)\varphi^t(s)
  +\varphi^{-1}(s) (d\exp\dot\psi(s))^t)(x)\\
 \nonumber &=&
 ( -\varphi^{-1}(s)
  d\exp[\psi(s),M]\tilde{\beta}_{\varphi(s)}
  + {d\over dt}\Big|_{t=0}\varphi^{-1}(s)(d\exp[\psi(s),M])^t)(x).\\
\end{eqnarray}
The last  term in \eqref{e:beta1} is
\begin{eqnarray}\label{e:beta2}
\nonumber  
\lefteqn{\frac{d}{dt}\Big|_{t=0}\left(\varphi^{-1}(s)(d\exp[\psi(s),M])^t\right)(x)}\\ 
\nonumber &=&
 {d\over
 dt}\Big|_{t=0}\left(\varphi^{-1}(s)\varphi(s)^t
 (\varphi(s)^t)^{-1}(d\exp[\psi(s),M])^t\right)(x)\\
\nonumber &=&
 {d\over dt}\Big|_{t=0}\left(\varphi^{-1}(s)\varphi(s)^t\right)\
 \left((\varphi(s)^t)^{-1}(d\exp[\psi(s),M])^t\right)\Big|_{t=0}(x)\\
\nonumber && +\left(\varphi^{-1}(s)\varphi(s)^t\right)\Big|_{t=0}\ \
 {d\over dt}\Big|_{t=0}\left((\varphi(s)^{-1})^{t}(d\exp[\psi(s),M])^t\right)(x)\\
\nonumber &=&
 \left(\tilde{\beta}_{
 \varphi(s)}\star(\varphi(s)^{-1}d\exp[\psi(s),M])\right)(x)
 +{d\over dt}\Big|_{t=0}\left((\varphi(s)^{-1}(d\exp[\psi(s),M]))^t\right)(x)\\
\nonumber &=&
 \left(\tilde{\beta}_{
 \varphi(s)}\star(\varphi(s)^{-1}d\exp[\psi(s),M])\right)(x)
 +|x|\lambda\left(\varphi(s)^{-1}(d\exp[\psi(s),M])\right)(x)\\
\nonumber  &=&
 \left(\tilde{\beta}_{
 \varphi(s)}\star(\varphi(s)^{-1}d\exp[\psi(s),M])\right)(x)
 +\left(\lambda(\varphi(s)^{-1}(d\exp[\psi(s),M]))\circ Y\right)(x)\nonumber
\end{eqnarray}
Substituting  back into \eqref{e:beta1} we get \eqref{e:t:phi}.
\end{proof}



\begin{corollary}\label{cor:beta}   Let $\varphi\in G^\Phi_\mathcal A$. Let
$\dot\psi(s)=[M,\psi(s)]$
 be the Lax pair from Theorem~\ref{t:8.2}
with  $\psi(0)=\psi = \log (\varphi)$.
 Let $\varphi(s)=\exp(\psi(s))$ and assume that $\varphi(s)\in
 G^\Phi_{\mathcal A}$ for all $s$.
Then
\begin{eqnarray*} {d\over ds}
 \beta_{\varphi(s)} &=&
 \mathrm{Ad}(\varphi(s)_+(0))\left([\tilde{\beta}_{\varphi(s)},\varphi^{-1}(s)\star
 d\exp[\log(\varphi(s)),M]]_+ \Big|_{\lambda=0}\right.\\
&&\qquad \left. +
 \mathrm{Res}\;\left((\varphi^{-1}(s)\star
 d\exp[\log(\varphi(s)),M])\circ Y\right)\right).
\end{eqnarray*}
\end{corollary}


\begin{proof} This follows from the previous Theorem and
 Lemma~\ref{lemma:6.6}.
\end{proof}

\begin{remark} In general, $\varphi\in G^\Phi_{\mathcal A}$ does not imply
 $\varphi(s)= \exp(\psi(s))\in G^\Phi_{\mathcal
A}$ for all $s$ (see Theorem
\ref{t:h3}).
A simple example with $\varphi(s)\in
G^\Phi_{\mathcal A}$ is given by a holomorphic $\varphi$  (i.e
$\varphi(x)\in\mathcal A_+$) with $-n+2m=0$. Indeed
$(\varphi^t)_-=\varepsilon$ as $\varphi^t$ is holomorphic, so
$(\varphi^t)_-$ does not depend on $t$. From the Taylor series of
the exponential,  $\exp(-s\log(\varphi))$ has only a
holomorphic part so $g_-(s)=\varepsilon$. Therefore the solutions $\psi(s)$ of the Lax
pair equation are constant, so $\varphi(s)=\varphi(0)\in
G^\Phi_{\mathcal A}$. 

\end{remark}

\section{The Lax pair flow and the renormalization group
flow}\label{s:rge}

The Lax pair flow lives on the Lie algebra $\mathfrak g_{\mathcal A}$ 
of infinitesimal characters, while
the beta character flow is on the Lie group $G_{\mathcal A}$ 
of characters.  Theorem~\ref{t:flow} and
Corollary~\ref{cor:beta} show that under the exponential map $\exp:{\mathfrak
  g}_{\mathcal A}\to G_{\mathcal A}$, 
the corresponding flow of beta characters
and $\beta$-functions are not in Lax pair form.  The main point of this
section is that the bijection $\tilde R^{-1}
:{\mathfrak   g}_{\mathcal A}\to G_{\mathcal A}$ of \cite{man} is much better
behaved: under $\tilde R^{-1}$, local characters remain local under the Lax pair flow 
(Theorem~\ref{t:7.9}), and the beta characters and the
$\beta$-functions satisfy Lax pair equations (Theorems~\ref{t:corr}).
In contrast, we give a rooted trees  example of the nonlocality of the Lax pair flow of
characters using the exponential map.  



\subsection{The pole order under the Lax pair flow}

 To begin, we investigate the dependence of the pole order of
the Lax pair  flow $L(t)$ on the pole order of the initial
condition $L_0$ and the Casimir function (e.g. the
 functions $\psi_{m,n}$). 
In the rooted trees case, the computations are considerably simplified 
using the normal coordinates of \cite{chr}, which we refer to for details.

Let $H$ be the Hopf algebra of rooted trees and $\mathcal{T}$ 
the set of trees. We choose a order on
$\mathcal{T}=\{t_i\}_{i\in\mathbb{N}}$, such that $deg(t_i)\leq
deg(t_j)$ for any $i<j$ and such that $h(t_i)\geq h(t_j)$ for any
trees $t_i,t_j$ with $deg(t_i)=deg(t_j)$ and $i<j$. Here $deg(t)$
is the number of vertices of $t$, and $h(t)$ is the height of
the tree $t$,  the length of the path from the root to the
deepest node in the tree.
 For example, we can choose
$$t_0=1_\mathcal{T}, \;\;\;\;
 t_1= \ta1,           \;\;\;\;
 t_2= \tb2,           \;\;\;\;
 t_3= \tc3,           \;\;\;\;
 t_4=\td31,          \;\;\;\;
 t_5=\te4,          \;\;\;\;
 t_6=\tf41,          \;\;\;\;
 t_7=\thj44,          \;\;\;\;
 t_8=\th43
$$

We recall that the $\exp:\mathfrak g_{\mathcal A}\to G_\mathcal{A}$
is bijective with inverse
 $\log:G_\mathcal{A}\to\mathfrak g_{\mathcal A}$  given by
 $$\log(\varphi)=\sum_{k=1}^\infty
 (-1)^{k-1}\frac{(\varphi-\varepsilon)^{k}}{k}.$$
Set $f_0= 1_\mathcal{T}$ and let $\{f_i\}_{i\in\mathbb{N^*}}$ be the
normal coordinates,  i.e.~$f_i$ is the
forest in $H$ satisfying
$$\log(\varphi)(t_i)=(\varphi-\varepsilon)(f_i),$$
for every character $\varphi$.
 For example,
$$
   f_1=\ta1\; , \;\;\;\; f_2=\tb2-\frac{1}{2}\ta1\ta1 \; ,\;\;\;\;
   f_3=\tc3-\ta1\tb2+\frac{1}{3}\ta1^3\; ,\;\;\;\;
   f_4=\td31-\ta1\tb2+\frac{1}{6} \ta1^3\;.
$$
$$
   f_5=\te4-\ta1\tc3-\frac{1}{2}\tb2^2+\ta1^2\tb2-\frac{1}{4}\ta1^4\; ,\;\;\;\;
   f_8=\th43-\frac{3}{2}\ta1\td31+\frac{1}{2}\ta1^2\tb2
   \;.
$$

For a ladder tree $t$, the forest $f$ given by
$\log(\varphi)(t )=(\varphi-\varepsilon)(f )$ for every character
$\varphi$,
 is a primitive element of the Hopf algebra $H$.

We identify $\varphi\in G_\mathcal{A}$ with
$\{\varphi(f_i)\}_{i\in\mathbb{N}}\in\mathcal{A}^\mathbb{N}$ and
 call the $\varphi(f_i)$ the $i$-component of $\varphi$.
Since $\varphi(f_0)=1$ for all $\varphi$, we drop the
$0$-component.

We use Sweedler's notation for the reduced coproduct
$\tilde\Delta(x)=x'\otimes x''$, where
$\tilde\Delta(x)=\Delta(x)-x\otimes 1_{\mathcal{T}}-  1_{\mathcal{T}}\otimes
x$. Notice that $\deg(x')+\deg(x'')=\deg(x)$ and
 $1\leq\deg(x'),\ \deg(x'')<\deg(x).$
For $x\neq 1_{\mathcal{T}}$ and
 $\tilde\Delta(x')=(x')'\otimes (x')''$, we have
\begin{eqnarray*}
 ((\varphi_1\varphi_2)\varphi_1^{-1})(x)
 &=& \langle
   (\varphi_1\varphi_2)\otimes\varphi_1^{-1},
    x\otimes1_{\mathcal{T}}
    +1_{\mathcal{T}}\otimes x
    + x'\otimes x''\rangle\\
   &=& (\varphi_1\varphi_2)(x)+\varphi_1^{-1}(x)
   +(\varphi_1\varphi_2)(x')\varphi_1^{-1}(x'')\\
&=&   \varphi_1(x)+\varphi_2(x)
    +\varphi_1(x')\varphi_2(x'')
    +\varphi_1^{-1} (x)+\big(\varphi_1(x')+\varphi_2(x')\\
&&\qquad
    +\varphi_1((x')')\varphi_2((x')'')\big)\varphi_1^{-1}(x'')\\
&=& \varphi_2(x)
    +\varphi_1(x')\varphi_2(x'')
    + \big(\varphi_1(x)+\varphi_1^{-1}(x)
    +\varphi_1(x')\varphi_1^{-1}(x'')\big)\\
&&\qquad 
    +\varphi_2(x' )\varphi_1^{-1}(x'')
    +\varphi_1((x')')\varphi_2((x')'') \varphi_1^{-1}(x'')\\
&=& \varphi_2(x)
    +\varphi_1(x')\varphi_2(x'')
    +\varphi_2(x' )\varphi_1^{-1}(x'')
    +\varphi_1((x')')\varphi_2((x')'') \varphi_1^{-1}(x'')
\end{eqnarray*}
Differentiating with respect to $\varphi_2$ and setting $L=\dot\varphi_2$
gives the adjoint representation:
\begin{eqnarray}\label{e:adjoint-repres}
    \mathrm{Ad}(\varphi_1)(L)(x)&=&L(x)
    +\varphi_1(x')L(x'')
    +L(x' )\varphi_1(S x'')\\
&&\qquad    +\varphi_1((x')')L((x')'') \varphi_1(Sx''),\nonumber
\end{eqnarray}
where $S$ is the antipode of the Hopf algebra.

\begin{theorem}\label{p:holo}
i) If the initial condition $L_0\in\mathfrak{g}_{\mathcal{A}}$ 
is holomorphic in $\lambda$, then the solution $L(t)={\Ad}(g_+(t))L_0$ of the Lax
pair equation is
holomorphic in $\lambda$.\\
ii) If $L_0\in\mathfrak{g}_\mathcal{A}$ has a pole of order $n$, then
 $L(t)={\Ad}(g_+(t))L_0$ has a pole of order at most
$n$.
\end{theorem}
\begin{proof} By  \eqref{e:adjoint-repres}, we have
\begin{eqnarray}\label{e:adjoint-repres2}
\mathrm{Ad}(g_+(t))(L_0)(x)&=& L_0(x)
    +g_+(t)(x')L_0(x'')
    +L_0(x')g_+(t)(Sx'')\\
&&\qquad     +g_+(t)((x')')L_0((x')'') g_+(t)(Sx'').\nonumber
\end{eqnarray}
Notice that $g_+(t)(x)$ is holomorphic for $x\in H$. If $L_0$ is
holomorphic, then every term of the right hand side of
\eqref{e:adjoint-repres2} is holomorphic, so
$\mathrm{Ad}(g_+(t))(L_0)$ is holomorphic. Since multiplication with
a holomorphic series cannot increase the pole order,
$L(t)$ cannot have a pole order greater than the pole order of
$L_0$.
\end{proof}

We can also use normal coordinates to measure the nontriviality of the Lax
pair flow.

\begin{theorem}\label{t:ladder}
If $f_i$ is a primitive element of $H$ (e.g.~$f_i$ corresponds to a
ladder tree), then  the $i$-component $L(t)(f_i)$ of the Lax pair
flow is does not depend on $t$.
\end{theorem}
\begin{proof}
It is shown in \cite{chr} that
 $\varphi^{-1}(f_i)=-\varphi(f_i),$ for every character $\varphi$ and $i\geq 1$.
Since $f_i$ is a primitive element, the inner automorphism
 $C_g:G_{\mathcal{A}}\to G_{\mathcal{A}}$, $C_g(h)=ghg^{-1}$,
 satisfies
 $C_g(h)(f_i)=g(f_i)+h(f_i)-g(f_i)=h(f_i).$
 Therefore
 $(Ad(g)L_0)(f_i)=(L_0)_i,$
 where $(L_0)_i$ is the i-component of $L_0$.
\end{proof}

Thus everything of interest in the Lax pair flow occurs off the
primitives, e.g.~the normal coordinate $f_4$ corresponding to
$\td31$ is the first component in the Hopf algebra on which the Lax
pair is nonconstant.

\subsection{The Lax pair flow, the RGE flow, and locality}

We now investigate whether the Lax pair flow can ever be identified with the RGF.
Some identification is necessary:  the RGF  $(\varphi^{t})_+(\lambda=0)$ lives
in the Lie group of characters $G_{\mathbb C}$, while the Lax pair flow
$L(t)$ lives in a Lie algebra $\mathfrak g_\mathcal{A}$ 
To match these flows, we can transfer the Lax pair 
flow to the Lie group level using either of the maps $\tilde
R^{-1}$ and $\exp$, namely by defining
\begin{equation}\label{phi-chi}
\varphi_t=\tilde R^{-1}(L(t))\ \ \ \text{and} \ \ \  \chi_t=\exp(L(t))
\end{equation}
and then setting $\lambda =0.$

The most naive hope would be that
$\varphi_t$ or $\chi_t$ coincide with the RGF $\varphi^t$, perhaps after
a rescaling of the parameter $t$.  We shall see that this fails even in the
trivial case.  As usual, we take $\mathcal A$ to be the algebra of
Laurent series.


\begin{proposition}
  On a  commutative, cocommutative, graded connected Hopf algebra $\mathcal H$, 
$\varphi_t\not=\varphi^t$ and $\chi_t\not=\varphi^t$.
\end{proposition}

\begin{proof}
If $\tilde R(\varphi^t)=L(t)$ then by Definition \ref{def:6.1}
   $$
     (\varphi^t\circ Y)(\Gamma)=(\varphi^t\star L(t))(\Gamma)
   $$
  for every $\Gamma\in \mathcal H$.
  For a primitive homogenous element
  $\Gamma\in {\mathcal H}_n$
  we get
\begin{equation}\label{7.4a}
      |\Gamma|e^{|\Gamma|t\lambda}\varphi(\Gamma)
      =e^{|\Gamma|t\lambda}\varphi(\Gamma)L(t)(1)
       +e^{0|\Gamma|\lambda}\varphi(\Gamma)L(t)(\Gamma).
\end{equation}
  Therefore
   $$L(t)(\Gamma)=|\Gamma|e^{|\Gamma|t\lambda}\varphi(\Gamma).$$
Since ${\mathcal H}$ is cocommutative, its Lie bracket is abelian.  Thus the
left hand side of (\ref{7.4a}) is constant in $t$, while the right hand side
is not.

  The same argument works for $\chi_t$ on $\mathcal H$. 
\end{proof}


In a positive direction, we will show that locality of characters is preserved
under the Lax pair flow, using the identification given by $\tilde R.$  This
indicates that $\tilde R$ is more useful than the exponential map.


Recall from \cite[Theorem IV.4.1]{man} that 
 $\lambda\tilde R:G_\mathcal{A}\to\mathfrak{g}_{\mathcal{A}}$
 restricts to a bijection from $G^\Phi_\mathcal{A}$ to
$\mathfrak{g}_{\mathcal{A}+}$, where $\mathfrak{g}_{\mathcal{A}+}$
is the set infinitesimal characters on $\mathcal H$ with values in $\mathcal
A_+$. 
In this sense, $\lambda \tilde R$ is better behaved than $\tilde R$, as the
 following locality result shows.

\begin{proposition}\label{t:7.9tau}
  For  a local character $\varphi\in G^\Phi_\mathcal{A}$, let $L(t)$ be the solution of the Lax
  pair equation (\ref{8:1}) with initial condition 
   $L_0=\lambda\tilde R(\varphi)$ and any $\mathrm{Ad}$-covariant
    function $f$.  Let
 $\tau_t$ be the flow of characters
  given by
    $$\tau_t=(\lambda\tilde R)^{-1}(L(t)).$$
Then $\tau_t$ is a local character for all $t$.
\end{proposition}

\begin{proof}
By \cite{ef-manchon},
$$\tilde R(\varphi\star\xi)=\tilde R(\xi)+\xi^{*-1}\star\tilde R(\varphi)\star
   \xi.
$$
 Taking $\xi=g_+(t)^{-1}$ and multiplying by $\lambda$, we get
$$\lambda\tilde R(\varphi\star g_+(t)^{-1})=\lambda\tilde R(g_+(t)^{-1})+
g_+(t)\star\lambda\tilde R(\varphi)\star
   g_+(t)^{-1}.
$$
 Since $\varphi\in G^\Phi_\mathcal{A}$ and $g(t)_+^{-1}$ is an element
 in $G_\mathcal{A}$ without polar part, by
  \cite[Lemma IV.4.3.]{man},
   $\varphi\star g(t)_+^{*-1}$
  is local.  Thus $\lambda\tilde R(\varphi\star g_+(t)^{-1})\in{\mathcal{A}_+}$.
We have $g(t)_+^{-1}\in G^\Phi_\mathcal{A}$, simply because
  $g(t)_+^{-1}$ does not have a polar part,
so $\lambda\tilde R (g(t)_+^{-1})$ is holomorphic. It follows that
  $$
\tau_t=(\lambda\tilde R)^{-1}({\Ad}(g_+(t))L_0)=(\lambda\tilde R)^{-1}(
g_+(t)\star L_0\star g_+(t)^{-1})\in G^\Phi_\mathcal{A}
  $$
\end{proof}


We can now show that locality of characters is preserved under the Lax pair
flow via the $\tilde R$ identification.

\begin{theorem}\label{t:7.9}
For a local character $\varphi\in G^\Phi_\mathcal{A}$,
  let $L(t)$ be the solution of the Lax pair equation \eqref{8:1}
  for any $\mathrm{Ad}$-covariant function $f$,
  with the initial condition $L_0=\tilde R(\varphi)$. Let
  $\varphi_t$ be the flow given by
    $$\varphi_t=\tilde R^{-1}(L(t)).$$
  Then $\varphi_t$ is a local character for all $t$.
\end{theorem}
\begin{proof}
We show that the flow $\tau_t$ constructed in the previous
 Proposition with the initial condition $L_0=\lambda\tilde
 R(\varphi)$,
 for the $\mathrm{Ad}$-covariant function
 $h:\mathfrak{g}_\mathcal{A}\to\mathfrak{g}_\mathcal{A}$ given
 by $h(L)=f(\lambda^{-1} L)$, is equal to the flow $\varphi_t$ constructed
 for the $\mathrm{Ad}$-covariant function $f$,  with the initial condition
 $L_0=\tilde
 R(\varphi)$.
We have
 $$\tau_t=(\lambda\tilde R)^{-1}\left(g_+(t)\star\lambda\tilde
 R(\varphi)\star g_+^{-1}(t)\right)=\tilde R^{-1}\left(g_+(t)\star \tilde
 R(\varphi)\star g_+^{-1}(t)\right),
$$
where $g_+(t)$ is given by the Birkhoff decomposition of
$$
\exp(-t f( \tilde R(\varphi)))= \exp(-t h(\lambda\tilde
R(\varphi))).$$ Therefore the two $g_+(t)$ involved in the
definitions of $\varphi_t$ and $\tau_t$ coincide, so
$\varphi_t=\tau_t$.
\end{proof}

In contrast to Theorem~\ref{t:flow}, it is immediate that 
the flow of beta characters associated to
$\tilde R$ is in Lax pair form.

\begin{lemma}\label{l:rminusoneversion}
For a local character $\varphi\in G^\Phi_\mathcal{A}$, let
$\varphi_t$ be the flow from Theorem 7.5. Then
\begin{eqnarray}\label{e:rminusoneversion}
\frac{d\tilde\beta_{\varphi_t}}{dt}=[\tilde\beta_{\varphi_t},M],
\end{eqnarray}
\end{lemma}
\begin{proof}
By Lemma 6.7, we get
 $\tilde\beta_{\varphi_t}=\lambda\tilde R(\varphi_t)=\lambda L(t)$. Then
$$\frac{d\tilde\beta_{\varphi_t}}{dt}= \frac{d(\lambda\tilde
  R(\varphi_t))}{dt}=
\frac{\lambda d
L(t)}{dt}=\lambda[L(t),M]=[\tilde\beta_{\varphi_t},M].$$
\end{proof}

The corresponding $\beta$-functions also satisfy a Lax pair equation.

\begin{theorem}\label{t:corr}  For a local character $\varphi\in G^\Phi_\mathcal{A}$,
let $L(t)$ be the Lax pair flow of Corollary 5.10 with  initial condition
$L_0=\tilde R(\varphi)$. Let $\varphi_t=\tilde R^{-1}(L(t))$. Then
\begin{itemize}
\item[(i)] for $-n+2m\geq 1$,  $\varphi_t = \varphi$ and hence
  $\beta_{\varphi_t}=\beta_{\varphi}$ for all $t$.
\item[(ii)] for $-n+2m\leq 0$,  $\beta_{\varphi_t}\in 
\mathfrak{g}_{\mathbb C}$ satisfies
$$\frac{d\beta_{\varphi_t}}{dt}=\big[\beta_{\varphi_t},
-\frac{d((\varphi_t)_+(0))}{dt}((\varphi_t)_+(0))^{-1} +
2\mathrm{Ad}((\varphi_t)_+(0))\big(\mathrm{Res}(\lambda^{-n+2m-2}\tilde\beta_{\varphi_t})\big)
\big].$$  
\end{itemize}
\end{theorem}

\begin{proof}
By Theorem 7.5, $\varphi_t$ are local characters, so by
\cite[Theorem IV.4.]{man}, $\tilde\beta_{\varphi_t}=\lambda
L(t)=\lambda \tilde R (\varphi_t)$ is holomorphic.

(i) If $-n+2m\geq 1$, then
 $\lambda^{-n+2m}L(t)$ is holomorphic, which implies
$$M=R(\lambda^{-n+2m}L(t))=\lambda^{-n+2m}L(t)=
 \lambda^{-n+2m-1}\tilde\beta_{\varphi_t}.$$
$L(t)$ satisfies the Lax pair equation
$$\frac{dL}{dt}=[L,M]=[L,\lambda^{-n+2m}L]=\lambda^{-n+2m}[L,L]=0.$$
Thus $L(t)=L_0$ for all $t$, which gives $\varphi_t=\tilde
R^{-1}(L(t))=\tilde R^{-1}(L_0)=\varphi$ for all $t$.

(ii) For $-n+2m\leq 0$, we have
\begin{eqnarray*}M &=&
  R(\lambda^{-n+2m}L(t))=\lambda^{-n+2m}L(t)-2P_-(\lambda^{-n+2m}L(t))\\
&=& 
\lambda^{-n+2m-1}\tilde\beta_{\varphi_t}-2P_-(\lambda^{-n+2m-1}\tilde\beta_{\varphi_t}).
\end{eqnarray*}
(\ref{e:rminusoneversion}) becomes
$$\frac{d\tilde\beta_{\varphi_t}}{dt}=
-2[\tilde\beta_{\varphi_t},P_-(\lambda^{-n+2m-1}\tilde\beta_{\varphi_t})]
=-2[P_+(\lambda^{-n+2m-1}\tilde\beta_{\varphi_t}),\lambda^{n-2m+1}
P_-(\lambda^{-n+2m-1}\tilde\beta_{\varphi_t})] 
.$$
Expand
 $\tilde\beta_{\varphi_t}$  as
 $$\tilde\beta_{\varphi_t}=\sum_{k=0}^{\infty} \tilde\beta_k(t)\lambda^k.$$
 Then
$$\frac{d\tilde\beta_{\varphi_t}}{dt}=-2\left[\sum_{k=n-2m+1}^\infty
\tilde\beta_{k}(t)\lambda^{k-n+2m-1},\sum_{j=0}^{n-2m}\tilde\beta_j(t)\lambda^j\right],
$$
and evaluating at $\lambda=0$ gives
\begin{eqnarray}
\frac{d\tilde\beta_0(t)}{dt}=-2[\tilde\beta_{n-2m+1}(t),\tilde\beta_0(t)]
=2[\tilde\beta_0(t),\tilde\beta_{n-2m+1}(t)].
\end{eqnarray}
Using the facts that Ad$(g)$ is a Lie algebra homomorphism and
$(d/dt){\rm Ad}(g(t))X$\\ 
$ = [(dg(t)/dt)g^{-1}(t), {\rm Ad}(g(t))X]$ (see the proof
of Theorem~\ref{t:7.9tau}), we get
\begin{eqnarray}
\frac{d\beta_{\varphi_t}}{dt}&=&
\frac{d}{dt}\left(\mathrm{Ad}(\varphi_t)( \tilde\beta_0(t)) \right)\nonumber\\
& =&
\left[\frac{d((\varphi_t)_+(0))}{dt}((\varphi_t)_+(0))^{-1},
 \mathrm{Ad}(\varphi_t)(\tilde\beta_0(t))\right]
+
\mathrm{Ad}((\varphi_t)_+(0))\left(\frac{d\tilde\beta_0(t)}{dt}\right)\\
&=&
\left[\frac{d((\varphi_t)_+(0))}{dt}((\varphi_t)_+(0))^{-1},\beta_{\varphi_t}\right]
+2\left[\mathrm{Ad}((\varphi_t)_+(0))\tilde\beta_0(t),
\mathrm{Ad}((\varphi_t)_+(0))\tilde\beta_{n-2m+1}(t)\right]\nonumber\\ 
&=&\left[\beta_{\varphi_t},-\frac{d((\varphi_t)_+(0))}{dt}((\varphi_t)_+(0))^{-1
}+2\mathrm{Ad}((\varphi_t)_+(0))\left(\mathrm{Res}(\lambda^{-n+2m-2}\tilde\beta_{\varphi_t})
\right)\right],\nonumber 
\end{eqnarray}
since
$\tilde\beta_{n-2m+1}(t)=\mathrm{Res}(\lambda^{-n+2m-2}\tilde\beta_{\varphi_t})$.
\end{proof}

Local characters satisfy the
 abstract Renormalized Group Equation \cite{0609035}, which we now recall.
For a local character $\varphi\in G^\Phi_{\mathcal A}$,  the renormalized
character is defined by
 $\varphi_{\mathrm{ren}}(t)=(\varphi^t)_+(\lambda=0)$.

\begin{theorem}\label{aRGE} For $\varphi\in G^\Phi_{\mathcal A}$,  the
 renormalized character
 $\varphi_{\mathrm{ren}}$ satisfies the abstract Renormalized Group
  Equation:
   $$\frac{\partial}{\partial t}
   \varphi_{\mathrm{ren}}(t)=\beta_\varphi\star\varphi_{\mathrm{ren}}(t).$$
\end{theorem}

Here our parameter $t$ corresponds to $e^t$ in \cite{0609035}.

In light of Theorem \ref{t:7.9},
we can ask for the relation between
$(\varphi_t)_{\mathrm{ren}}(s)$ and
$\varphi_{\mathrm{ren}}(s)$ corresponding to $\varphi_t$ and
$\varphi$.
In \S9, we consider a toy model character on a
Hopf algebra of rooted trees and show that these renormalized characters
differ.

We can also show that for certain initial conditions, the flow $\tau_t$ is constant.

\begin{proposition}
If $\varphi\in G^\Phi_\mathcal{A}$ and $\varphi_+=\varepsilon$ (i.e.
$\varphi$ has only a pole part), then the flow $\tau_t$ of Proposition \ref{t:7.9tau}
for    the $\mathrm{Ad}$-covariant function 
   $f(L)=\lambda^{-n+2m}L$ has $\tau_t=\varphi$ for all $t$. 
\end{proposition}
\begin{proof}
If we show that either $g_\pm(t)=\varepsilon$, 
then
 $$\tau_t=(\lambda\tilde R)^{-1}\big(g_\pm(t)\star\lambda\tilde R
 (\varphi)\star g_\pm(t)^{-1}\big)=(\lambda\tilde R)^{-1}\big(\varepsilon\star\lambda\tilde R
 (\varphi)\star\varepsilon^{-1}\big)=\varphi.
 $$
 $g_\pm(t)$ are given by the Birkhoff decomposition of
\begin{eqnarray}\label{e:expforX}
g(t)=\exp(-2t\lambda^{-n+2m}L_0)=\sum_{k=0}^\infty
\frac{(-2t\lambda^{-n+2m})^k L_0^{k}}{k!},
\end{eqnarray}
 where
  $L_0=\lambda\tilde R(\varphi)\in\mathfrak{g}_{\mathbb C}$
 \cite[Theorem IV.4.4]{man}. We split the problem into
  two cases depending on the sign of $-m+2n$.
  If $-m+2n\geq 0$, then $g(t)(x)\in\mathcal A_+$ for any $x$, which
  implies $g_-(t)=\varepsilon$.
  Similarly, if $-m+2n< 0$, then $g(t)(x)\in\mathcal A_-$ for any $x$, which
  implies  $g_+(t)=\varepsilon$. Notice that the right hand side of
 (\ref{e:expforX}) is  a finite sum, namely up to
  $k=\mathrm{deg}(x)$
  when evaluated  on $x\in \mathcal H$.
\end{proof}

We next study the locality of the flow $\chi_t$ defined in (\ref{phi-chi}) for
the usual Lax pair flow $L(t).$ .
Thus for an initial character $\varphi$ and $L_0 = \log(\varphi)$,
$$\chi_t =\exp(g_+(t)\star L_0\star g_+(t)^{-1})=g_+(t)\star
\exp(L_0)\star g_+(t)^{-1}=g_+(t)\star \varphi\star g_+(t)^{-1},$$
with $\chi_0=\varphi$.  As before, in normal coordinates $\chi_t$ is trivial
on primitives.

  \begin{lemma}\label{l:constant}
If $\varphi\in G^\Phi_{\mathcal A} $ and $f_i$ is a primitive
element, then $ \chi_t(f_i)$ does not depend on $t$.
\end{lemma}
\begin{proof}
$\chi_t(f_i)=g_+(t)(f_i)+\varphi(f_i)-g_+(t)(f_i)=\varphi(f_i)$.
\end{proof}

We now present some calculations showing the interplay between the Lax pair
flow and locality.

For the first example,
we construct a nontrivial Hopf subalgebra on which $\chi_t$ is local.
 Let $\mathcal H^{2}$ be the Hopf subalgebra generated by
the following trees
$$t_0=1_\mathcal{T}, \;\;\;\;
 t_1= \ta1,           \;\;\;\;
 t_2= \tb2,           \;\;\;\;
 t_4=\td31,
$$
together with any set of  ladder trees.  The first
normal coordinate of $\chi_t$ to depend on $t$ is $f_4$, corresponding
to $\td31$. Let $G^2_\mathcal{A}$ be the group of characters
associated to the  $\mathcal H^2$.

\begin{proposition}\label{t:7.9a}
For $\varphi\in G^{2\, \Phi}_{\mathcal A}$, let $\chi_t$ be the flow of characters
on $\mathcal H^2$  given by
    $$\chi_t=\exp(L(t)),$$
   where $L(t)$ is the solution of the Lax pair equation (\ref{8:1}) 
for any $\mathrm{Ad}$-covariant function
 with the initial condition
   $L_0=\log(\varphi)$.
  Then $\chi_t$ is local for all $t$.
\end{proposition}

\begin{proof}
Let $\pi$ denote the projection to the pole part of a Laurent
series. By \cite{chr}, $\tilde\Delta(f_4)=f_1\otimes
f_2-f_2\otimes f_1$, so
  $$(\chi_t^s)_-(f_4)=-\pi\big(\chi_t^s(f_4)+(\chi_t^s)_-(f_1)\chi_t^s(f_2)
     -(\chi_t^s)_-(f_2)\chi_t^s(f_1)\big).$$
Subtracting from this equation the  corresponding equation for $t=0$,
and remembering that $\chi_t(f_1)$ and $\chi_t(f_2)$ do not
depend on $t$ by  Lemma \ref{l:constant}, we get
$$(\chi_t^s)_-(f_4)=(\varphi^s)_-(f_4)-\pi(\chi_t^s(f_4)-\varphi^s(f_4)).
$$
We have
\begin{eqnarray*}\label{e:pole}
  \pi(\chi_t^s(f_4)-\chi^s(f_4))
    &=& \pi
        \big(
            e^{3\lambda s}
              (
                -2g_+(t)(f_2)\varphi(f_1)
                +2g_+(t)(f_1)\varphi(f_2)
              )
        \big)\\
    &=& \pi
      \big(
        - 2g_+(t)(f_2)\pi(e^{3\lambda s}\varphi(f_1))
        + 2g_+(t)(f_1)\pi(e^{3\lambda s}\varphi(f_2))
      \big).
\end{eqnarray*}
Since $\varphi\in G^{2\, \Phi}_\mathcal{A}$, both
 $\pi(e^{s\lambda}\varphi(f_1))=-(\varphi^s)_-(f_1)$ and
 $\pi(e^{2s\lambda}\varphi(f_2))=-(\varphi^s)_-(f_2)$
are independent of $s$. By rescaling $s$, 
$\pi(e^{3s\lambda}\varphi(f_1))$ and
$\pi(e^{3s\lambda}\varphi(f_2))$
are independent of $s$. Therefore
$\pi(\chi_t^s(f_4)-\varphi^s(f_4))$ is independent of $s$, which finishes
 the proof.
\end{proof}

\begin{remark}
We can apply the previous proposition to  the Hopf subalgebra of Feynman
diagram generated by the empty graph and the graphs
\begin{eqnarray*}
    A_1=  \p\ ,\ A_2= \pdp\ ,\ A_3= \pdpdp\ ,\
    A_4= \pddpp\ ,\ A_5= \pdpdpdp
\end{eqnarray*}
and with $\varphi$ the Feynman rules character. 
The  characters $\chi_t$ restricted to this Hopf algebra are all local.
\end{remark}

To investigate how $\chi_t$ fails to be local on a
 non-ladder tree with a larger number of vertices, we 
consider  the Hopf subalgebra $\mathcal H^3$ generated by
$$t_0=1_\mathcal{T}, \;\;\;\;
 t_1= \ta1,           \;\;\;\;
 t_2= \tb2,           \;\;\;\;
 t_3=\tc3,          \;\;\;\;
 t_4=\td31,          \;\;\;\;
t_6=\tf41,          \;\;\;\;
 t_7=\thj44,          \;\;\;\;
 t_8=\th43
$$
together with any set of ladder trees.
 Let $f_i$ be
the corresponding normal coordinates. 

The next lemma gives the pole order of a
local character $\varphi\in G^\Phi_{\mathcal A}$ on primitives.

\begin{lemma}\label{lemma:7.2}
 If $\varphi$ is a local character and
 $f_i$ is primitive, then both $\varphi(f_i)$ and
  $L_0(f_i)=\log(\varphi)(f_i)$ have a pole of
 order at most one.
\end{lemma}
\begin{proof}
 If ${\rm deg}(f_i)=d$ and $\varphi=\sum_{k=-n}^\infty \varphi_k\lambda^k$
 is the
 Laurent expansion of $\varphi$, then
 $$(\varphi^s)_-(f_i)=-\pi(e^{sd\lambda}\varphi(f_i))=
   -\pi(
\varphi_{-n}(f_i)\lambda^{-n}+(\varphi_{-n}(f_i)sd+\varphi_{-n+1}(f_i))\lambda^{-n+1}
+o(\lambda^{-n+2})) 
 $$
 If $\varphi$ has a pole, then  $\varphi^s_- = \varphi_-$ implies
 $-n+1=0$.
\end{proof}

\begin{proposition}\label{t:h3}
   Let $\varphi$ be a local character on
 $\mathcal H^3$  and let $\chi_t$ be the flow of characters
  given by $$\chi_t=\exp(L(t)),$$
   where $L(t)$ is the solution of the Lax pair equation 
(\ref{8:1})  with the initial condition
   $L_0=\log(\varphi)$. 
Then $\chi_t$ is local on $\mathcal H^3$ for all $t$ if and only if
either
    \begin{eqnarray}\label{e:condition:chi}
\varphi_-(f_1)=0\ \ \text{or} \ \
3\varphi_-(f_1)(\varphi_+(f_2)\big|_{\lambda=0})=\varphi_-(f_2)(\varphi_+(f_1)\big|_{\lambda=0}).
\end{eqnarray}
\end{proposition}

The point is that (\ref{e:condition:chi}) is unlikely to hold.

\begin{proof}

By \cite{chr},
$$\tilde\Delta(f_8)=\frac{3}{2}f_1\otimes f_4-\frac{3}{2}f_4\otimes f_1
              -\frac{1}{2}f_1\otimes f_1 f_2-\frac{1}{2}f_1f_2\otimes f_1
              +\frac{1}{2}f_1f_1\otimes f_2+\frac{1}{2}f_2\otimes f_1f_1.
$$

Since
 $(\chi^s_t)_-(f_4)$ does not depend on $s$
(Prop.~\ref{t:7.9a}) and since $\chi_t(f_1)$ and $\chi_t(f_2)$ do
not depend on $t$,  after  cancelations of terms
involving only the primitives $f_1$ and $f_2$, we get
 \begin{eqnarray}\label{e:tf8}
\lefteqn{
  (\chi^s_t)_-(f_8)-(\varphi^s)_-(f_8)}\\
&=&
   -\pi
     \Big(\chi^s_t(f_8)-\varphi^s(f_8)\\
   &&\qquad      +\frac{3}{2}\varphi_-(f_1)(\chi^s_t(f_4)-\varphi^s(f_4))
         -\frac{3}{2}\varphi^s(f_1)((\chi_t)_-(f_4)-\varphi_-(f_4))
     \Big)\nonumber
\end{eqnarray}

By  Lemma \ref{lemma:7.2}, $\varphi(f_1)$ and $\varphi(f_2)$ have
poles of order at most one. Set
$$\varphi(f_1)=\sum_{k=-1}^\infty a_k\lambda^k\ \ \ \text{and}\ \ \
\varphi(f_2)=\sum_{k=-1}^\infty b_k\lambda^k.$$

From the proof of Proposition \ref{t:7.9a},
\begin{eqnarray}\label{e:chi1}
  \chi_t^s(f_4)-\varphi^s(f_4))
    =
                e^{3\lambda s}
              (
                -2g_\pm(t)(f_2)\varphi(f_1)
                +2g_\pm(t)(f_1)\varphi(f_2)
              )
\end{eqnarray}
We have
$g_-(t)(f_1)=\exp(-2t\lambda^{-n+2m}L_0)_-(f_1)=-\pi(-2t\varphi(f_1))=2ta_{-1}\lambda^{-1}$,
and\\
 $g_+(t)(f_1)=-2t(a_0+a_1\lambda+o(\lambda^2))$.
  Similarly
$g_-(t)(f_2)=2tb_{-1}\lambda^{-1}$.
(\ref{e:chi1}) becomes
 $\chi_t^s(f_4)-\varphi^s(f_4))=4te^{3\lambda
    s}\big((-b_{-1}a_0+a_{-1}b_0)\lambda^{-1}+o(\lambda^0)\big)$, 
 which implies
\begin{eqnarray*} \lefteqn{
  \pi\big(\frac{3}{2}\varphi_-(f_1)(\chi^s_t(f_4)-\varphi^s(f_4))
         -\frac{3}{2}\varphi^s(f_1)((\chi_t)_-(f_4)-\varphi_-(f_4))\big)}\\
&=&
  \frac{3}{2}\pi\big((-a_{-1}\lambda^{-1}e^{3s\lambda}(4t((-b_{-1}a_0+a_{-1}b_0)
\lambda^{-1}+o(\lambda^0)))\big)\\ 
&&\qquad
-\frac{3}{2}\pi\big(e^{s\lambda}(a_{-1}\lambda^{-1}+o(\lambda^0))(4t)(-1)(-b_{-1}a_0
    +a_{-1}b_0)\lambda^{-1}\big)\\ 
&=& -12sta_{-1}(-b_{-1}a_0+a_{-1}b_0)\lambda^{-1}+P(\lambda^{-1}),
\end{eqnarray*}
where $P(\lambda^{-1})$ is some polynomial in $\lambda^{-1}$ independent
of  $s$. We get
 \begin{eqnarray}\label{e:tf8s}
\lefteqn{  \pi
     \Big(\chi^s_t(f_8)-\varphi^s(f_8)
     \Big)}\\
    & =&
     \pi
       \big(
          e^{4s\lambda}
             (
                (  -3g_+(t)(f_1)g_+(t)(f_2)-3g_+(t)(f_3)
                )\varphi(f_1)
             )
       \big)\\
       &&\qquad +3\pi
       \big(e^{4s\lambda}
                (  g_+(t)(f_1)  )^2\varphi(f_2)
                \big)
        +\pi
       \big(3e^{4s\lambda}
                 g_+(t)(f_1)\varphi(f_4)
            \big)\nonumber
\end{eqnarray}
Since $\varphi(f_1)$ and $\varphi(f_2)$ have poles of order at most
one,  the first two terms of the right hand side of
 (\ref{e:tf8s}) do not depend on $s$.
%
Since $\varphi\in G^{3\, \Phi}_\mathcal{A}$, 
$$\varphi^s_-(f_4)=-\pi\big(e^{3s\lambda}\varphi(f_4)+\varphi_-(f_1)e^{2s\lambda}\varphi(f_2)
-\varphi_-(f_2)e^{s\lambda}\varphi(f_1))\big)
$$
is independent  of $s$, and so
$\pi(e^{3s\lambda}\varphi(f_4)-a_{-1}b_{-1}s\lambda^{-1})$ is also independent of
$s$. By rescaling $s$, 
$\pi(3e^{4s\lambda}\varphi(f_4)-4a_{-1}b_{-1}s\lambda^{-1})$ does
not depend on $s$. In conclusion,  the  terms independent of $s$
in $-(\chi^s_t)_-(f_8)+(\varphi^s)_-(f_8)$ are
$$
-12sta_{-1}(-b_{-1}a_0+a_{-1}b_0)\lambda^{-1} +
4a_{-1}b_{-1}s\lambda^{-1}(-2ta_0).
$$
Therefore $\chi^s_t(f_8)$ is independent of $s$ if and only if
either $a_{-1}=0$ or $3a_{-1}b_0=a_0b_{-1}$. Similar computations
hold for the   normal coordinates $f_6$ and $f_7$.
\end{proof}

\begin{remark}
 The choice $-n+2m=0$ in the Proposition is just for the sake
 of concreteness. A more detailed analysis reveals the
 following:
\begin{itemize}
\item[-]
 If $-n+2m\geq 1$ then $\chi_t$ is local  on $\mathcal H^3$
 without any additional conditions.
 Indeed, in this case  $L_0(f_1)$ and
 $L_0(f_2)$ are holomorphic and thus $g_-(t)(f_1)=g_-(f_2)=0$. which
 implies that $\chi^s_t(f_4)-L_0(f_4)=0$ for every $t$. By
 (\ref{e:tf8}) we get that $(\chi_t^s)_-(f_8)$ does not depend on
 $s$. Similar statements hold for  $f_6$ and $f_7$.
\item[-]
 If $-n+2m=-1$, the situation is similar to Proposition
 \ref{t:h3}, namely $\chi_t$ is local on $\mathcal H^3$ if
 and only if
$$\varphi_-(f_1)(\varphi_+(f_2)\big|_{\lambda=0})=\varphi_-(f_2)(\varphi_+(f_1)\big|_{\lambda=0})$$
and either
 $$\varphi_-(f_1)=0\ \ \text{or} \ \
 3\varphi_-(f_1)(\frac{\partial\varphi_+(f_2)}{\partial\lambda}\big|_{\lambda=0})
 =\varphi_-(f_2)(\frac{\partial\varphi_+(f_1)}{\partial\lambda}\big|_{\lambda=0}).
 $$
\end{itemize}
For $-n+2m\in\Z^+$, 
 the flows
 $\chi_t$ and $L(t)$ gain locality, in the sense that they become constant 
on  larger Hopf subalgebras as $-n+2m$ increases.  Indeed,
 $\chi_t$ and $L(t)$ are constant on the Hopf algebra generated by
 the primitives (e.g. the normal coordinates associated to ladder
 trees. 
In contrast, if we decrease
 $-n+2m<0$,  we preserve locality
 only when an increasing number of conditions are
  fulfilled.
\end{remark}

\subsection{The Lax pair flow of the $\beta$-function}

Recall from \S6 that the beta characters for the exponentiated Lax pair flow
 $\exp(L(t))$ do not
themselves satisfy a Lax pair equation.  In the next theorem, we reverse this
 procedure by taking a Lax pair flow $L(t)$ starting at the $\beta$-function of a
 character, and then producing characters $\xi_t$ whose $\beta$-functions are
 $L(t)\big|_{\lambda = 0}.$


\begin{theorem}\label{t:7.14}
   Let $\varphi\in G^\Phi_{\mathcal A}$ and let
   $L(t)$ be the flow given by Theorem~\ref{t:8.2}
with the initial condition
$L_0=\beta_\varphi.$ Let
   $\xi_t=(\lambda \tilde{R})^{-1}(L(t)\big|_{\lambda=0})$.
   Then $\xi_t$ is local for all  $t$.  The $\beta$-function of $\xi_t$
   satisfies
$$\beta_{\xi_t} = L(t)\big|_{\lambda =0}.$$
 Moreover,
     $\xi(0)=  \varphi_-$.
\end{theorem}
\begin{proof}
Since $L_0=\beta_\varphi$ is scalar valued, by Theorem \ref{p:holo}
   $L(t)=\mathrm{Ad}(g_+(t))(L_0)$
   is  holomorphic. Therefore. $L(t)$ can be evaluated at
   $\lambda=0$.
   $L(t)\big|_{\lambda=0}$ is also scalar valued, so by
   \cite[Theorem IV.4.4]{man}, $\xi_t\in G^\Phi_{\mathcal A-}$, the
   set of local characters taking values in
   $\lambda^{-1}\mathbb C[\lambda^{-1}]$.
In
   particular, $\xi_t$ is local.  By Lemma \ref{lemma:6.6}, $\tilde
   \beta_{\xi_t} = (\lambda \tilde R)(\xi_t) = L_t\big|_{\lambda = 0}$, so
   $\tilde\beta_{\xi_t}$ must be constant in $\lambda.$ This implies 
$$\beta_{\xi_t} = \tilde\beta_{\xi_t}\big|_{\lambda =0} = L(t)\big|_{\lambda =0}.$$
It follows from the  Connes-Kreimer scattering formula
   \cite{ck2,man} that $\xi(0)=\varphi_-$.
\end{proof}

\section{ Worked examples}\label{worked-example}


In this section we give some explicit computations on two Hopf algebras which
illustrate results in previous sections.
We first consider the Hopf
algebra $\mathcal H^1$ generated by the following trees:
$$t_0=1_\mathcal{T}, \;\;\;\;
 t_1= \ta1,           \;\;\;\;
 t_2= \tb2,           \;\;\;\;
 t_4=\td31,          \;\;\;\;
 t_8=\th43,
$$
and the regularized toy model character
$\varphi=\varphi(q,\mu,\lambda)$ (see \cite{krei99,krde99})
given on trees by
$$\varphi(T)(q,\mu,\lambda)=(q/\mu)^{-\lambda\mathrm{deg(T)}}\prod_{v}
B_{w(T_v)}(\lambda).$$
 Here the product is taken over all vertices $v$ of the tree $T$,
  $w(T_v)$ is the number of vertices of the subtree $T_v$ of $T$ which has
 $v$ as a root, and $B_j(\lambda)=B(j\lambda,1-j\lambda)$ for
 $j\in\mathbb N^*$, with  $B$ the Euler beta function. 
Referring to
 \cite{ef-manchon}, $q$ is interpreted as a dimensional external
 parameter, and $\mu$ is the 't Hooft mass. 
$\varphi$ has enough similarity of realistic QFT calculations to be worth considering
 \cite{bk99,krde99,krei99,krei00}.
Set
 $b=q/\mu$ and $a=\log(b)$. Thus terms in $a$ (or $\log(q^2/\mu^2) = 2a$ as in
e.g.~\cite{ky}) are the leading log terms in the various expansions.
We have
 $\varphi(\ta1)=b^{-\lambda}B_1(\lambda)$,\
 $\varphi(\tb2)=b^{-2\lambda}B_2(\lambda)B_1(\lambda)$,\
 $\varphi(\td31)=b^{-3\lambda}B_3(\lambda)B_1(\lambda)^2$,
 \
 $\varphi(\th43)=b^{-4\lambda}B_4(\lambda)B_1(\lambda)^3$, etc.
In the normal coordinates $f_i$, the Laurent series of
$\varphi(f_i)$ (cf. \cite{chr}) are given by
$$\varphi(f_1)=\frac{1}{\lambda}-a+o(\lambda),\ \ \ \ \
 \varphi(f_2)=\frac{\pi^2}{4}+o(\lambda),
$$
$$
\varphi(f_4)=\frac{7\pi^2}{36 \lambda} -\frac{7\pi^2a}{12} +
o(\lambda), \ \ \
  \varphi(f_8)=\frac{\pi^2}{12 \lambda^2}
-\frac{\pi^2a}{3\lambda} + o(\lambda^0).
$$

The character $\varphi$ is local, and the Lax pair flow on $\mathcal H^1$
$\varphi_t=\tilde R^{-1}(L(t))$ 
as in Corollary \ref{tc:8.2},
with $L(t)=Ad(g_\pm(t))L_0$,
$-m+2n=0$, and initial condition $L_0=\tilde R(\varphi)$ is
given by
$$\varphi_t(f_1)=\varphi(f_1),\ \ \ \ \varphi_t(f_2)=\varphi(f_2),\ \
\ \varphi_t(f_4)=\frac{\pi^2 (7 + 24 t)}{36 \lambda}
-\frac{\pi^2}{12} (7 + 16 t)a + o(\lambda),
$$
\begin{eqnarray*}\varphi_t(f_8)&=&\frac{\pi^2 (1 + 6 t)}{12 \lambda^2} - \frac{\pi^2(2 + 15 t + 9
t^2) a}{6 \lambda}\\ 
&&\qquad  + \frac{\pi^2}{144} (\pi^2 (83 + 288 t + 126 t^2)
+ 12 (8 + 51 t + 27 t^2) a^2)+ o(\lambda).
\end{eqnarray*}
The flow $L(t)$ has poles of
order at most one:
$$L(t)(f_1)=\frac{1}{\lambda}-a+o(\lambda),\ \ \
 L(t)(f_2)=\frac{\pi^2}{2}+o(\lambda),$$
$$L(t)(f_4)=
   \frac{\frac{\pi^2}{3} + 2 \pi^2 t}
   {\lambda}
   + (-\pi^2  - 4 \pi^2 t) a +o(\lambda),
$$
$$L(t)(f_8)=\frac{-2 (\pi^2 t (2 + 3 t)a)}{\lambda}
  +\frac{\pi^2}{6}
       (\pi^2 (7 + 37 t + 21 t^2) + 6 t (8 + 9 t) a^2)+o(\lambda).$$
This confirms that  $\lambda L(t)=(\lambda\tilde
R)(\varphi_t)=\tilde\beta_{\varphi_t}$ is holomorphic, which implies
that $\varphi_t$ is local  on $\mathcal H^1$. It  can be explicitly checked
that $(\varphi_t^s)_-$ does not depend on
$s$:
$$
  (\varphi_t^s)_-(f_1)=-\frac{1}{z},  \ \ \ \ \ \
  (\varphi_t^s)_-(f_2)=0, \ \ \ \ \ \ \
 (\varphi_t^s)_-(f_4)=\frac{\pi^2(1-12t)}{18\lambda},
$$
$$
 (\varphi_t^s)_-(f_8)=-\frac{\pi^2(1-12t)}{24 \lambda^2}+\frac{3\pi^2t(1+t)a}{2\lambda}.
$$
The Connes-Kreimer $\beta$-functions 
$\beta_{\varphi_t}=-{\rm Res}(\varphi_t)_-\circ Y$ are:
$$\beta_{\varphi_t}(f_1)=1,\ \ \ \beta_{\varphi_t}(f_2)=0,\ \ \
\beta_{\varphi_t}(f_4)=-\frac{\pi^2}{6} + 2 \pi^2 t, \ \ \ \
\beta_{\varphi_t}(f_8)=-6 \pi^2 t (1 + t)a.
$$
The associated RGFs
$(\varphi_t)_{\mathrm{ren}}(s)=(\varphi_t^s)_+\big|_{\lambda=0}$,
 which all satisfy the abstract RGE, are
   $$ (\varphi_t)_{ren}(s)(f_1)=(s - a), \ \ \ \  \ (\varphi_t)_{ren}(s)(f_2)=\frac{\pi^2}{4},
   $$
   $$
    (\varphi_t)_{ren}(s)(f_4)=
      \frac{\pi^2}{12}(s + 24 s t - (1 + 16 t)a),
   $$
$$(\varphi_t)_{ren}(s)(f_8)=\frac{\pi^2}{96}
    (12 s^2 + \pi^2 (11 + 136 t + 84 t^2) - 24 s (1 + 20 t + 24 t^2)a + 12 (1 + 22 t + 18 t^2)
    a^2).$$

Thus in this model 
there is a polynomial dependence in $t$ of the leading log terms in each of
$L(t), \beta_{\varphi_t}$ and $(\varphi_t)_{\rm ren}$, although the first
diagram with a nonzero leading log term differs.  By the recursion formula in
\cite[(26)]{ky}, the next to ... leading log terms in the Green's functions 
will then also depend polynomially on $t$.  We conjecture that this polynomial
dependence extends to the Feynman rules character on the full Hopf algebra.

We emphasize that the Renormalized Group Flows
$(\varphi_t)_{\mathrm{ren}}$ and $\varphi_{\mathrm{ren}}(s)=
(\varphi_t)_{\mathrm{ren}}(s)\big|_{t=0}$ of the characters
$\varphi_t$ and $\varphi$ are different. While $\varphi_t(f_1),
\varphi_t(f_2)$ are independent of $t$, we have
\begin{equation}\label{8:1new}
\varphi_t(f_3)=\varphi(f_3)-4t(\varphi_-(f_1)\varphi(f_2)-\varphi_-(f_2)\varphi(f_1).
\end{equation}
We have used $\varphi_t=\tilde
R^{-1}(\mathrm{Ad}(g_-(t))\tilde R(\varphi))$ for this calculation, since it
is easier to extract the pole part of a Laurent series than the holomorphic
part, but we could also use
$\varphi_t=\tilde
R^{-1}(\mathrm{Ad}(g_+(t))\tilde R(\varphi))$.  In this case, we get
\begin{equation}\label{8:2}
\varphi_t(f_3)=\varphi(f_3)-4t(\varphi_+(f_1)\varphi(f_2)-\varphi_+(f_2)\varphi(f_1).
\end{equation}
As a check, we verify that (\ref{8:1new}) and (\ref{8:2}) are equal.
 Let $\pi$ denote the projection onto the pole part of a
Laurent series. Then
\begin{eqnarray*} \varphi_+(f_1)\varphi(f_2)-\varphi_+(f_2)\varphi(f_1)
&=& (\varphi(f_1)-\pi(\varphi(f_1))\varphi(f_2)
(\varphi(f_2)-\pi(\varphi(f_2))\varphi(f_1)\\
&=& -\pi(\varphi(f_1)\varphi(f_2) -(-\pi(\varphi(f_2)))\varphi(f_1)\\
&=&\varphi_-(f_1)\varphi(f_2)-\varphi_-(f_2)\varphi(f_1).
\end{eqnarray*}

The computations for the character $\chi_t$ in (\ref{phi-chi}) associated to the
toy model character $\varphi$ with  $-n+2m=0$ give 
 $$(\chi_t^s)_-(f_1)=-\frac{1}{\lambda},\ \ \ \ \ (\chi_t^s)_-(f_1)=0, \
 \ \ \ \
 (\chi_t^s)_-(f_4)=-\frac{-\frac{\pi^2}{18}+\pi^2t}{\lambda},$$
$$(\chi_t^s)_-(f_8)=\frac{\pi^2}{24\lambda^2}+\frac{\pi^2t(18s+(5+18t)a)}{6\lambda}.$$
In agreement with Theorem \ref{t:7.9a},  $(\chi_t^s)_-$ is independent of $s$ when evaluated on
$f_1$, $f_2$ and $f_4$.
However, $(\chi_t^s)_-(f_8)$ depends on $s$, so
 $\chi_t$ is not local. We confirm that the necessary
 condition (\ref{e:condition:chi}) for locality in Theorem \ref{t:h3}
 does not hold. Indeed, $\varphi_-(f_1)=1/z\not= 0$ and
 $3(-1/z)(\pi^2)\not=0\cdot a$. The $\beta$-function on $f_1$, $f_2$, $f_4$ is given by
 $$\beta_{\chi_t}(f_1)=1, \ \ \ \ \ \beta_{\chi_t}(f_2)=0,\ \ \
\ \ \
 \beta_{\chi_t}(f_4)=-\frac{\pi^2}{6}+3\pi^2t.$$
 The
 renormalized character  $(\chi_t)_{\mathrm{ren}}(s)$ is given by
 $$(\chi_t)_{\mathrm{ren}}(s)(f_1)=s-a, \ \ \
 (\chi_t)_{\mathrm{ren}}(s)(f_2)=\pi^2/4, \ \ \
(\chi_t)_{\mathrm{ren}}(s)(f_4)=\frac{\pi^2}{12}(s+36st-(1+24t)a),$$
and satisfies the abstract RGE.


Let $\mathcal H^2$ be the Hopf subalgebra generated by the trees
$$t_0=1_\mathcal{T}, \;\;\;\;
 t_1= \ta1,           \;\;\;\;
 t_2= \tb2,           \;\;\;\;
 t_3= \tc3,           \;\;\;\;
 t_4=\td31,          \;\;\;\;
 t_5=\te4
 .
$$
%
%
%
For $T\in \{t_1,\ldots , t_{5}\}$, let $Z_T$ be the
corresponding infinitesimal character.  The Lie
algebra $\mathfrak g_2$ of scalar valued infinitesimal characters of $\mathcal H^2$
is generated by
$Z_{t_1},\ldots,\ Z_{t_{5}}$. 
Let $G_1$ be the scalar valued  character group 
 of $\mathcal H^2$, and 
let $G_0$ be the semi-direct product
   $G_1\rtimes{\mathbb C}$ given by
   $$(g,t)\cdot (g',t')=(g\cdot\theta_t( g'),t+t'),$$ where
   $\theta_t(g)(T)=e^{t\mathrm{deg}(T)}g(T)$ homogenous $T$.
Define a new variable $Z_0$ with $[Z_0,Z_{t_i}] = \mathrm{deg}(t_i)Z_{t_i},$
   so formally
   $Z_0=\frac{d}{d\theta}$.
 The Lie algebra $\mathfrak g_0$ of $G_0$ is
   generated by $Z_0,Z_{t_1},\ldots,\ Z_{t_{5}}$.

 The conditions a) and b) in Definition
\ref{bialg} of a Lie bialgebra can be written in a basis as a system
of quadratic equations. We can solve this system explicitly, e.g.
via Mathematica. It turns out that
there are $43$ families of Lie bialgebra structures $\gamma$
   on $\mathfrak g_0$.
In more detail, the system of quadratic equations
involves 90 variables.  Mathematica gives
   1 solution with 82 linear relations (and so 8 degrees of freedom),
   7 solutions with 83 linear relations,
   16 solutions with 84 linear relations,
   13 solutions with 85 linear relations,
   5 solutions with 86 linear relations, and
   1 solution with 87 linear relations.


  To any Lax equation with a spectral parameter, one can associate a
  spectral curve and  study its algebro-geometric properties
  (see \cite{sts}).
  In our case,  we consider the adjoint representation $\mathrm{ad}:\delta\to
  {\mathfrak gl}(\delta)$ and the induced adjoint representation of the loop
  algebra. The spectral curve is given by the characteristic equation
  of $\mathrm{ad}(L\lambda)$:
   $\Gamma_0=\{ (\lambda,\nu)\in{\mathbb C}-\{0\}\times{\mathbb C}\ | \
  \det(\mathrm{ad}(L(\lambda)-\nu {\mathrm{ Id} }))=0\}$.

  The theory of the spectral curve and its Jacobian usually assumes that the
  spectral curve is irreducible.
  For all 43 families of Lie bialgebra structures on $\delta$, on the associated
  Lie algebra
   $\mathrm{ad}(\delta)$ all eigenvalues of the characteristic
  equation
  are  zero, and the zero eigenspace is nine dimensional.  
The spectral curve
  itself is the union of degree one curves.
  Thus each  irreducible component has a trivial Jacobian, and the
  spectral curve theory breaks down. The integrability of
these Lax pair equation remains open for future investigations.

\begin{acknowledgments}
Gabriel Baditoiu would like to thank the Max-Planck-Institute for
Mathematics, Bonn and the Erwin Schr\"odinger International
Institute for Mathematical Physics for the hospitality.  Steven Rosenberg
would also like to thank ESI and the Australian National University.
\end{acknowledgments}

\bibliographystyle{amsplain}
\bibliography{paper-new2}
\end{document}